\documentclass[12pt]{article}
\usepackage{epsfig}
\usepackage{amsmath}
\usepackage{hhline}
\usepackage{amssymb}
\usepackage{times}
\usepackage{cite}
\usepackage{graphicx}
\usepackage{ifpdf}
\ifpdf
   \usepackage{epstopdf}
   \usepackage{pdfsync}
\fi

\newlength{\dinwidth}
\newlength{\dinmargin}
\begin{document}
\newcommand{\pom}{{I\!\!P}}
\newcommand{\reg}{{I\!\!R}}
\def\gsim{\,\lower.25ex\hbox{$\scriptstyle\sim$}\kern-1.30ex%
\raise 0.55ex\hbox{$\scriptstyle >$}\,}
\def\lsim{\,\lower.25ex\hbox{$\scriptstyle\sim$}\kern-1.30ex%
\raise 0.55ex\hbox{$\scriptstyle <$}\,}
\newcommand{\trm}{m_{\perp}}
\newcommand{\trp}{p_{\perp}}
\newcommand{\trmm}{m_{\perp}^2}
\newcommand{\trpp}{p_{\perp}^2}
\newcommand{\alp}{\alpha_s}
\newcommand{\alps}{\alpha_s}
\newcommand{\sqrts}{$\sqrt{s}$}
\newcommand{\PT}{p_{\perp}}
\newcommand{\JPSI}{J/\psi}
\newcommand{\PO}{I\!\!P}
\newcommand{\xbj}{x}
\newcommand{\xpom}{x_{\PO}}
\newcommand{\dgr}{^\circ}
\newcommand{\gev}{\,\mbox{GeV}}
\newcommand{\GeV}{\mathrm{GeV}}
\newcommand{\GeVS}{\rm GeV^2}
\newcommand{\xp}{x_p}
\newcommand{\xpi}{x_\pi}
\newcommand{\xg}{x_\gamma}
\newcommand{\xgj}{x_\gamma^{jet}}
\newcommand{\xpj}{x_p^{jet}}
\newcommand{\xpij}{x_\pi^{jet}}
\renewcommand{\deg}{^\circ}
\newcommand{\qsq}{\ensuremath{Q^2} }
\newcommand{\gevsq}{\ensuremath{\mathrm{GeV}^2} }
\newcommand{\et}{\ensuremath{E_t^*} }
\newcommand{\rap}{\ensuremath{\eta^*} }
\newcommand{\gp}{\ensuremath{\gamma^*}p }
\newcommand{\der}{{\mathrm d}}
\def\Journal#1#2#3#4{{#1} {\bf #2}, #4 (#3)}
\def\NCA{\em Nuovo Cimento}
\def\NIM{\em Nucl. Instrum. Methods}
\def\NIMA{{\em Nucl. Instrum. Methods} {\bf A}}
\def\NPB{{\em Nucl. Phys.}   {\bf B}}
\def\PLB{{\em Phys. Lett.}   {\bf B}}
\def\PRL{\em Phys. Rev. Lett.}
\def\PRD{{\em Phys. Rev.}    {\bf D}}
\def\PR{{\em Phys. Rev.}    }
\def\ZPC{{\em Z. Phys.}      {\bf C}}
\def\ZP{{\em Z. Phys.}      }
\def\EJC{{\em Eur. Phys. J.} {\bf C}}
\def\EJA{{\em Eur. Phys. J.} {\bf A}}
\def\CPC{\em Comp. Phys. Commun.}

\begin{titlepage}

\begin{flushleft}
{\tt DESY 14-035    \hfill    ISSN 0418-9833} \\
{\tt March 2014}                              \\
\end{flushleft}

\vspace*{10mm}

\rm

\begin{center}
\begin{Large}
{\boldmath \bf
   Measurement of Feynman-$x$ Spectra of Photons and Neutrons
   in the Very Forward Direction in  Deep-Inelastic Scattering at HERA
}

\vspace{2cm}
\rm
H1 Collaboration

\end{Large}
\end{center}

\vspace{2cm}

\begin{abstract}
\noindent
\rm 
Measurements of normalised cross sections 
for the production of photons and neutrons 
at very small angles with respect to the proton beam direction
in deep-inelastic $ep$ scattering at HERA are presented 
as a function of the Feynman variable $x_F$ and 
of the centre-of-mass energy of the
virtual photon-proton system $W$.
The data are taken with the H1 detector in the years 2006 and 2007 and
correspond to an integrated luminosity of $131~\mathrm{pb}^{-1}$.
The measurement is restricted to photons and neutrons in the
pseudorapidity range  $\eta>7.9$ and covers the range of
negative four momentum transfer squared at the positron vertex
$6<Q^2<100$~GeV$^2$, of inelasticity $0.05<y<0.6$ and of
$70<W<245~$GeV.
To test the Feynman scaling hypothesis the
$W$ dependence of the $x_F$ dependent cross sections is investigated.
Predictions of deep-inelastic scattering models and of models for hadronic 
interactions of high energy cosmic rays are compared to the measured cross
sections.

\end{abstract}

\vspace*{10mm}

\it \normalsize
\begin{center} submitted to {\it Eur.Phys.J.C}
\end{center}

\end{titlepage}
\begin{flushleft}

\noindent
V.~Andreev$^{22}$,             
A.~Baghdasaryan$^{34}$,        
K.~Begzsuren$^{31}$,           
A.~Belousov$^{22}$,            
P.~Belov$^{10}$,               
V.~Boudry$^{25}$,              
G.~Brandt$^{46}$,              
M.~Brinkmann$^{10}$,           
V.~Brisson$^{24}$,             
D.~Britzger$^{10}$
% ...


\bigskip\noindent{\it
% ...
$ ^{2}$ School of Physics and Astronomy, University of Birmingham,
          Birmingham, UK$^{ b}$ \\
% ...
$ ^{10}$ DESY, Hamburg, Germany \\
% ...
$ ^{22}$ Lebedev Physical Institute, Moscow, Russia \\
% ...
$ ^{24}$ LAL, Universit\'e Paris-Sud, CNRS/IN2P3, Orsay, France \\
 $ ^{25}$ LLR, Ecole Polytechnique, CNRS/IN2P3, Palaiseau, France \\
% ...
$ ^{31}$ Institute of Physics and Technology of the Mongolian
          Academy of Sciences, Ulaanbaatar, Mongolia \\
% ...
$ ^{34}$ Yerevan Physics Institute, Yerevan, Armenia \\
% ...
$ ^{46}$ Department of Physics, Oxford University,
          Oxford, UK$^{ b}$ \\

\bigskip \noindent
% ...
$ ^b$ Supported by the UK Science and Technology Facilities Council,
      and formerly by the UK Particle Physics and
      Astronomy Research Council \\
% ...
}

\end{flushleft}

\newpage


\section{Introduction}

\noindent
Measurements of particle production at very small polar
angles with respect to the proton beam direction ({forward direction})
in positron-proton collisions are
important inputs for the theoretical
understanding of proton fragmentation mechanisms.
Forward particle measurements are also valuable
for high energy cosmic ray experiments, as they 
provide important new constraints for high energy air shower models
\cite{Engel:1998hf,Bunyatyan:2009zz}.

The H1 and ZEUS experiments at the $ep$ collider HERA have
studied the production of forward baryons (protons and neutrons) and photons, 
which carry  a large fraction of the longitudinal
momentum of the incoming proton
\cite{Adloff:1998yg,Chekanov:2002pf,Chekanov:2007tv,Chekanov:2008tn,Aaron:2010ab,Aaron:2011pe}.
These analyses have demonstrated that models of deep-inelastic scattering (DIS) 
are able to reproduce the forward baryon measurements if 
contributions from different production mechanisms are considered,
such as string fragmentation, pion exchange, diffractive dissociation and elastic
scattering of the proton~\cite{Chekanov:2008tn,Aaron:2010ab}.
The forward photon production rate, however, is overestimated by the models
by $50$ to $70$\%~\cite{Aaron:2011pe}.
The measurements  
also confirm the 
hypothesis of limiting fragmentation~\cite{Benecke:1969sh,Chou:1994dh},
according to which, in the high-energy limit, the cross
section for the inclusive production of particles in the target
fragmentation region is independent of the incident projectile energy.  

Measurements in the DIS regime provide a possibility to investigate
the process at different centre-of-mass (CM) energies
of the  virtual photon-proton system, $W$, within the same experiment.
The studies of the energy dependence of particle production 
allow a test of the Feynman scaling \cite{Feynman:1969ej} hypothesis,
according to which particle production 
is expected to show a scaling behaviour, 
i.e. independence of the CM energy in terms of the Feynman-$x$ variable,
$x_F=2p_{||}^*/W$.
Here  $p_{||}^*$ is the longitudinal momentum of the particle
in the virtual photon-proton CM frame with respect to the direction 
of the beam proton.
In several previous measurements Feynman scaling was found to be 
violated in the fragmentation process 
in the central rapidity region 
\cite{Arnison:1982ed,Arnison:1982rm,Banner:1984wh,Bernard:1985kh,Alner:1986xu,Alner:1987wb,Alexopoulos:1988na,Abe:1989td,Aaron:2007ds,Abramowicz:2010rz}.
On the contrary,  no  sizable violation of
Feynman scaling has been observed in the  target fragmentation region
in  $pp$ and $p\bar{p}$ collisions
by comparing the  $\pi^0$ production cross sections
at the SPS collider \cite{Pare:1989mr} with
$\pi^{\pm}$ 
measurements at the
ISR \cite{Albrow:1973kj,Albrow:1973ee,Capiluppi:1973fz,Capiluppi:1974rt}.
However, these conclusions are debated \cite{Ohsawa:1992dj}
and the scarcity of other experimental forward particle
production data motivates
further studies of forward particle production.

\begin{figure}[h]
\epsfig{file=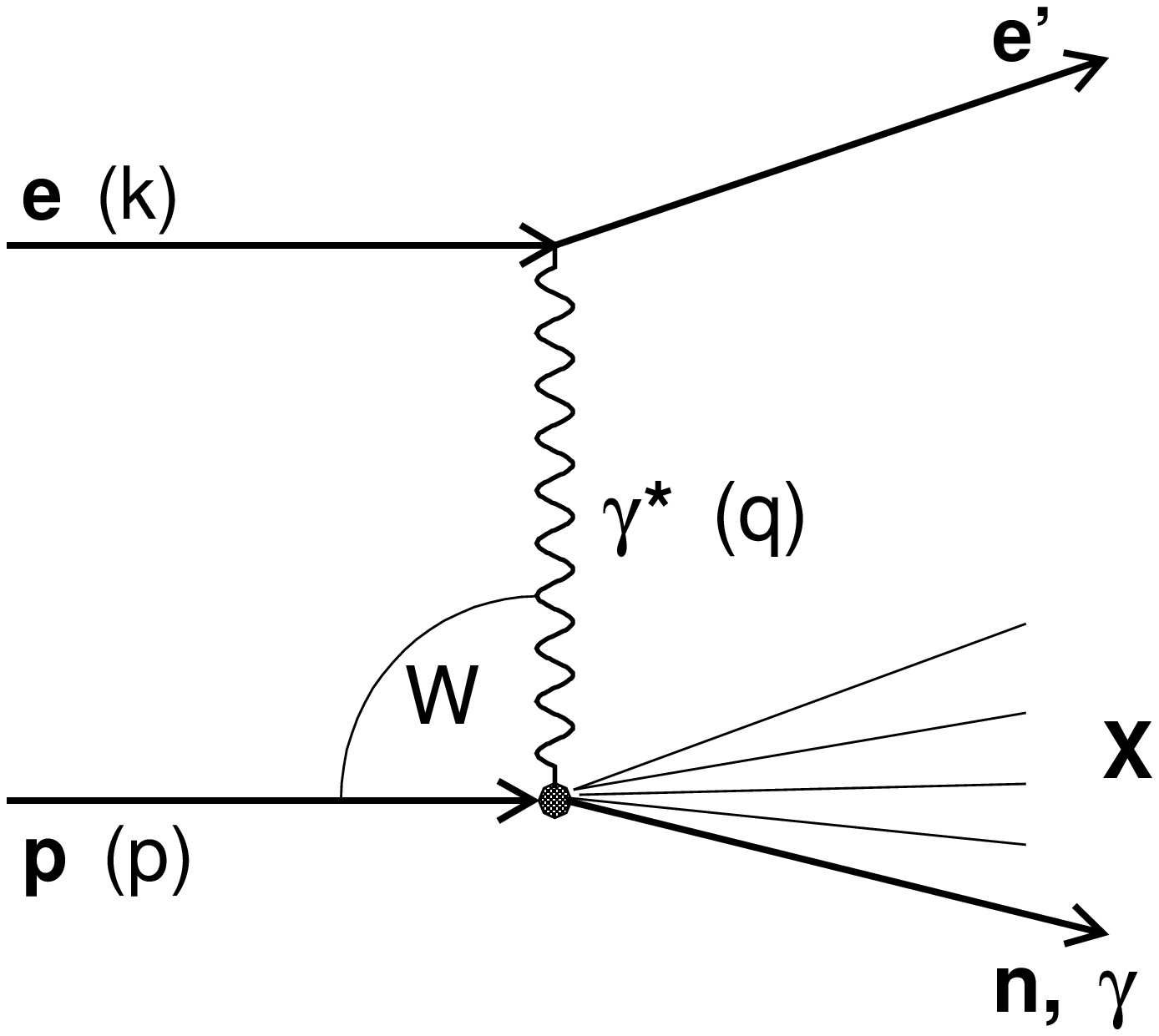,height=62mm}
\hspace*{15mm}
\epsfig{file=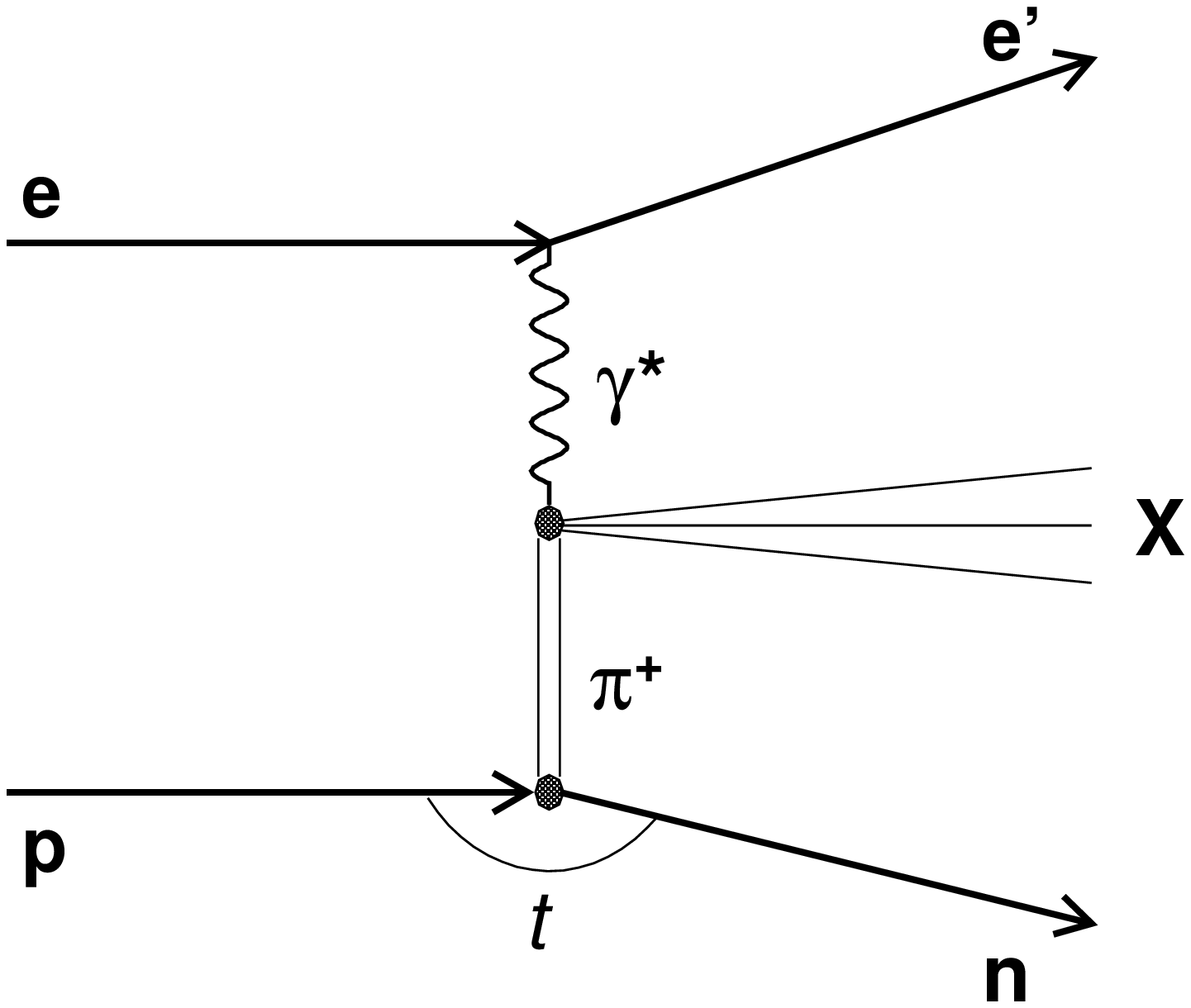,height=62mm}
\caption{ (a) Generic diagram for forward photon or neutron production 
$ep\rightarrow e'\gamma X$, $ep\rightarrow e'nX$ in deep-inelastic scattering.
(b) Diagram of forward neutron production via pion exchange.}

\vspace*{-79mm}
\hspace*{25mm}{\large \bf (a)} \hspace*{78mm} {\large \bf (b)}

\vspace*{77mm}  
\label{diagram1}
\end{figure}

In this paper the production of forward neutrons and photons in
DIS is studied 
as a function of $x_F$\footnote{In the kinematic 
range of this measurement the variable $x_F$ is numerically almost 
equal to the longitudinal momentum fraction $x_L$ used in the 
previous publications \cite{Adloff:1998yg,Chekanov:2002pf,Chekanov:2007tv,Chekanov:2008tn,Aaron:2010ab,Aaron:2011pe}.
There, $x_L$ was defined as $x_L=E_{n,\gamma}/E_p$,
where $E_p$, $E_{n}$ and $E_{\gamma}$ are the energies of
the proton beam, the forward neutron and the forward photon in the
laboratory frame, respectively.} and $W$.
This is the first direct experimental test of Feynman scaling
for  photons and neutrons produced in the very forward direction.
Predictions from different DIS and different cosmic ray (CR) 
hadronic interaction 
Monte Carlo (MC) models are compared to the results.
The simultaneous measurement of  forward neutrons and photons 
provides a useful input for
MC model development also because of  the respective different 
production mechanisms:
forward photons almost exclusively originate from
decays of neutral mesons produced in the fragmentation of the proton remnant 
(Figure~\ref{diagram1}a),
while forward neutrons  are produced also
via a colour singlet exchange process
(Figure~\ref{diagram1}b).

The neutrons and photons studied here are produced at polar angles
below $0.75$~mrad and are measured in the Forward Neutron 
Calorimeter (FNC) of the H1 detector.
The data used in this analysis were collected with the H1 detector at
HERA in the years 2006 and 2007 and correspond to an integrated
luminosity of $131~ \rm pb^{-1}$. During this period 
HERA collided positrons and protons with energies of 
$E_e=27.6~\GeV$ and $E_p=920~\GeV$, respectively, corresponding to a
centre-of-mass energy of $\sqrt{s}=319~\GeV$.

\section{Experimental Procedure and Data Analysis}
\subsection{H1 main detector}

A detailed description of the H1 detector can be found elsewhere
\cite{Abt:1996hi,Abt:1996xv,Appuhn:1996na,Pitzl:2000wz,Andrieu:1993kh,Laycock:2012xg}.
Only the detector components relevant to this analysis are briefly
described here.
 The origin of the right-handed
 H1 coordinate system is the nominal $e^+p$ interaction point. 
 The direction of the proton beam defines  the positive $z$--axis; 
 the polar angle $\theta$ is measured with respect to  this axis. 
 Transverse momenta are measured in the $x$--$y$ plane.
 The pseudorapidity is defined by  $\eta =  -\ln{[\tan(\theta/2)]}$ 
 and is measured in the laboratory frame.

The 
interaction region is surrounded by a two-layer silicon
strip detector and large concentric drift chambers,
operated inside a $1.16$~T solenoidal magnetic field.
Charged particle momenta are measured  in the
angular range $15\deg<\theta<165\deg$.
The forward tracking detector is used to supplement track reconstruction 
in the region $7\deg<\theta<30\deg$
and improves the hadronic final state 
reconstruction of forward going low momentum particles.
The tracking system is surrounded by a finely segmented
liquid argon (LAr) calorimeter, which covers the polar angle range
\mbox{$4\deg<\theta<154\deg$} with full azimuthal acceptance.  The LAr
calorimeter consists of an electromagnetic section with lead absorber
and a hadronic section with steel absorber.  The total depth of the
LAr calorimeter ranges from $4.5$ to $8$ hadronic interaction lengths $\lambda$.
 The absolute electromagnetic energy scale is
 known with a precision of $1\%$, while the absolute hadronic energy
 scale is known for the present data with a precision of $4\%$.

The backward region ($153\deg<\theta<177.8\deg$) is covered by a
lead/scintillating-fibre calorimeter called the SpaCal;
its main purpose is the detection of scattered positrons.  The polar angle of
the positron is measured with a precision of $1$~mrad.  The energy
resolution for
positrons is $\sigma(E)/E\approx
7.1\%/\sqrt{E[\GeV]}\oplus 1\%$ \cite{Nicholls:1995di} and the energy 
scale uncertainty is
less than $1\%$.  The hadronic energy scale in the SpaCal is known with
a precision of $7\%$.

 The luminosity is determined from the rate of the elastic QED Compton process 
 with the electron and the photon detected in the SpaCal calorimeter, 
 and the rate of DIS events measured in the SpaCal calorimeter 
 \cite{Aaron:2012kn}.

\subsection{Forward detector for neutral particles}
\label{sec:fncdet}

Neutral particles produced at very small polar angles with respect to the proton
beam direction can be detected in the FNC, 
which is situated 
at a polar angle of $0\deg$ at $106$~m from the interaction point.
A detailed description
of the detector is given in \cite{Aaron:2010ab,Aaron:2011pe}.  
The FNC is a lead--scintillator sandwich calorimeter. It consists of two 
longitudinal sections: the Preshower Calorimeter 
with a  length corresponding to about $60$ radiation lengths
or $1.6 \lambda$
 and the
Main Calorimeter with a total length of $8.9 \lambda$.
The acceptance of the FNC is defined by the aperture of the HERA
beam-line magnets and is limited to scattering angles of
$\theta\lsim 0.8$~mrad with approximately $30\%$ azimuthal coverage. 

The longitudinal segmentation of the FNC allows an efficient
discrimination of photons from neutrons.  
Photons are absorbed completely in the Preshower Calorimeter, while
neutrons  have a significant fraction of their energy 
deposited in the Main Calorimeter.
Therefore, energy deposits in the FNC, which are contained in the
Preshower Calorimeter
with no energy deposits above the noise level in the Main Calorimeter,
are classified as electromagnetic clusters.
According to the Monte Carlo simulation about $98\%$ of all reconstructed 
photon and neutron candidates originate from  generated photons and neutrons,
respectively. 
Due to the relatively large size of the FNC readout modules in
combination with the small  geometrical acceptance window, 
two or more particles entering the FNC are reconstructed as 
a single cluster.
In the MC simulation about $1\%$ of all
hadronic clusters in the FNC associated with neutrons
are overlapped with  a photon, which was scattered
within the FNC acceptance together with the neutron.
At lower energies the electromagnetic clusters reconstructed in the FNC
mainly originate from single photons. At higher measured energies there is 
a significant contribution from two photons, with the 
fraction of two-photon events increasing from $0.5\%$ at $100$~GeV to 
$10\%$ at about $450$~GeV and to $80\%$ at $900$~GeV.
The two photons typically originate from the decay of the same meson.

The absolute electromagnetic and hadronic energy scales of the FNC are known to 
$5\%$~\cite{Aaron:2011pe} and $2\%$~\cite{Aaron:2010ab} precision, respectively.
The energy resolution of the FNC calorimeter for electromagnetic
showers is $\sigma(E)/E \approx 20\%/\sqrt{E~[\rm GeV]} \oplus 2\%$
and for hadronic showers $\sigma(E)/E \approx 63\%/\sqrt{E~[\rm GeV]} \oplus 3\%$,
as determined in test beam measurements~\cite{Uraev}.
The spatial resolution is
$\sigma(x,y)\approx 10\rm cm/\sqrt{E~[\rm GeV]} \oplus 0.6~\rm cm$ for 
hadronic showers starting in the Main Calorimeter.
A better spatial resolution of about $2\rm~ mm$ is achieved for the
electromagnetic showers and for those hadronic showers 
which start in the Preshower Calorimeter.

\subsection{Cross Section Definition}

The kinematics of semi-inclusive forward photon and neutron production are shown
in Figure~\ref{diagram1}a, where the four-vectors of the incoming
and outgoing particles and of the exchanged virtual photon $\gamma^*$
are indicated.
This measurement is restricted to the DIS kinematic range, determined by the 
photon virtuality $6<Q^2<100~\GeV^2$ and inelasticity $0.05<y<0.6$.
They are defined as
\begin{equation}
Q^2=-q^2\; , \hspace{3em}  
y=\frac{p \cdot q}{p \cdot k}\;,
\end{equation}
where $p$, $k$ and $q$ are the four-momenta of the incident proton,
the incident positron and the virtual photon, respectively.
The  CM  energy of the virtual photon-proton system, $W$, is
related to $Q^2$ and $y$ as $W \approx \sqrt{ys-Q^2}$, 
where $s$ is the squared
total CM energy of the positron-proton system.
The present analysis is restricted to the range $70<W<245~\rm GeV$.

 The analysis is performed in the pseudorapidity range $\eta>7.9$
 for forward neutrons and  photons.
 The pseudorapidity range $\eta>7.9$ corresponds to polar
 angles $\theta<0.75$~mrad.
 In the virtual photon-proton CM frame
 the neutron transverse momentum $p_T^*$ and the neutron
 $x_F$
 are restricted to the 
 range $0<p_T^*<0.6~\rm GeV$ and $0.1<x_F<0.94$, respectively.
 For the forward photons measurement $p_T^*$ and $x_F$  
 are defined for the most energetic photon in the pseudorapidity
 range  $\eta>7.9$ and are  restricted to the 
 range $0<p_T^*<0.4~\rm GeV$ and $0.1<x_F<0.7$.  
 The requirement that $x_F$ is below $0.7$ for photons 
 ensures that the 
 electromagnetic clusters reconstructed in the FNC mainly  
 originate from single photons, according to MC predictions.

 The kinematic phase space of the measurements is
 summarised in Table~\ref{tab:selection}.
  Cross sections of neutrons and photons produced in the forward
  direction, normalised to the inclusive DIS cross section, 
  $ 1/\sigma_{DIS}~{\rm d}\sigma/{\rm d}x_F$, 
  are
  determined differentially in $x_F$ in three ranges of $W$.
  In addition, the cross section ratios integrated over $x_F$,
  $ \sigma^{\gamma,n}_{DIS}/\sigma_{ DIS}$, 
  are measured  as a function of $W$.

\begin{table}[h]
        \centering
        \renewcommand{\arraystretch}{1.50}
  \begin{tabular}{| l | l |}
                        \hline 
        \multicolumn{2}{|c|}{\bf NC DIS Selection}  \\ \hline
        \multicolumn{2}{|c|}{$6 < Q^2 < 100$~GeV$^2$}  \\
        \multicolumn{2}{|c|}{$0.05 < y < 0.6$} \\ 
        \multicolumn{2}{|c|}{$70 < W < 245$~GeV} \\ \hline
        {\bf Forward photons} & {\bf Forward neutrons} \\ \hline
        $\eta>7.9$ & $\eta>7.9$ \\
        $0.1<x_F<0.7$ & $0.1<x_F<0.94$ \\
        $0<p_T^*<0.4$~GeV & $0<p_T^*<0.6$~GeV \\ \hline
        \multicolumn{2}{|c|}{\bf \boldmath $W$ ranges for cross sections 
        {\large $ \frac{1}{\sigma_{DIS}}\frac{{\rm d}\sigma}{{\rm d}x_F}$}}  \\
        \multicolumn{2}{|c|}{$~70 < W < 130$~GeV} \\ 
        \multicolumn{2}{|c|}{$130 < W < 190$~GeV} \\ 
        \multicolumn{2}{|c|}{$190 < W < 245$~GeV} \\ \hline
  \end{tabular}
   \vspace{\baselineskip} 
        \caption{Definition of the kinematic phase space of the measurements.}
  \label{tab:selection}
\end{table}

\subsection{Event selection }

The  data selection and analysis procedures are similar to those 
described in previous publications using the FNC \cite{Aaron:2010ab,Aaron:2011pe}.
The data sample of this analysis was collected using
 triggers which require the scattered positron to be measured in the SpaCal.
 The trigger efficiency is about 96\% for the analysis phase space
 as determined from data using an independently triggered data sample.
The selection of DIS events is based on the identification
of the scattered positron as the most energetic, isolated compact calorimetric 
deposit
in the SpaCal  with an
energy $E_e'>11~\GeV$ and a polar angle $156\deg<\theta_e'<175\deg$.
The \mbox{$z$-coordinate} of the primary event vertex is required to be
within $\pm 35$~cm of the nominal interaction point.
The hadronic final state is reconstructed using an energy flow algorithm 
which combines charged particles measured in the trackers with information 
from the SpaCal and LAr calorimeters~\cite{Peez:2003zd,Hellwig:2004yp}.
To suppress events with hard initial state  QED
radiation, as well as
events originating  from  non-$ep$ interactions,
the quantity $\sum{E - p_z}$, summed over
all reconstructed particles including the positron, is required to lie
between  $35~\GeV$ and $70~\GeV$.
This cut also efficiently removes events from photoproduction processes, 
where the
positron is scattered into the backward beam-pipe and a particle from
the hadronic final state fakes the positron signature in the SpaCal.
The kinematic variables $Q^2$ and $y$ are reconstructed using a technique
which optimises the resolution throughout the
measured $y$ range, exploiting information from both the scattered
positron and the hadronic final state \cite{Adloff:1997sc}.
Events are restricted to the range
$6<Q^2<100~\GeV^2$ and $0.05<y<0.6$.
The DIS data sample contains about 9.3 million events.

A subsample of events containing forward photons or neutrons is 
selected by requiring 
either an electromagnetic or a hadronic cluster in
the FNC with a pseudorapidity above $7.9$ and an energy above $92~\rm GeV$. 
The data sample, called `FNC sample' in the following,
contains about $83 ,000$ events 
with photons and $230,000$ events with neutrons.

\subsection{Monte Carlo simulations and corrections to the data}
\label{sec:mc}

Monte Carlo simulations are used to correct the data for the
effects of detector acceptance, inefficiencies, QED
radiation from the positron and migrations between
measurement bins
due to the finite detector resolution.
All generated events are passed through
a GEANT3~\cite{Brun:1978fy} based simulation of the H1 apparatus and
are subject to the same reconstruction and analysis chain as the data.

The DJANGOH~\cite{Charchula:1994kf}
program is used to generate 
inclusive DIS events.  It is based on leading order
electroweak cross sections and takes into account QCD effects up to
order $\alpha_s$.
Higher order QCD effects are simulated using
leading log parton showers as implemented in LEPTO ~\cite{Ingelman:1996mq},
or using the Colour Dipole Model (CDM) as implemented in ARIADNE
\cite{Lonnblad:1992tz}.  Subsequent hadronisation effects are modelled
using the Lund string fragmentation model as implemented in JETSET
\cite{Andersson:1983ia,Sjostrand:1995iq}.  
Higher order electroweak processes are
simulated using an interface to HERACLES~\cite{Kwiatkowski:1990es}.
The LEPTO program optionally includes 
the simulation of soft colour interactions
(SCI)~\cite{Edin:1995gi}, in which the production of
diffraction-like configurations is enhanced via non-perturbative
colour rearrangements between the outgoing partons.
The SCI option in LEPTO is used for the simulation of forward photons.
For the DJANGOH MC simulations the H1PDF~2009
parameterisation~\cite{Aaron:2009kv} of the parton distributions in the
proton is used.  In the following, the DJANGOH predictions based on LEPTO
and ARIADNE are denoted as LEPTO and CDM, respectively.
 In all DJANGOH simulations  forward particles originate exclusively
 from the
 hadronisation of the proton remnant and forward photons 
 are therefore mainly produced from the decay of $\pi^0$ mesons.

RAPGAP~\cite{Jung:1993gf} is a general purpose event generator for 
inclusive and diffractive $ep$ interactions.  
Higher order QCD effects are simulated
using parton showers and the final state hadrons are
obtained via the Lund string model.  As in DJANGOH higher order electroweak
processes are simulated using an
interface to HERACLES~\cite{Kwiatkowski:1990es}.
In the version denoted below as RAPGAP-$\pi$, the program
simulates exclusively the scattering of virtual or real photons
off an exchanged pion (Figure~\ref{diagram1}b). 
In this model the cross section for $ep$ scattering to the final state 
$nX$ takes the form
\begin{equation}
  \der\sigma (e p\rightarrow e'nX) =f_{\pi^+/p}(x_L,t)\cdot
  \der\sigma(e\pi^+\rightarrow e'X).
\label{crsec}
\end{equation}
Here $x_L$ is the longitudinal 
momentum fraction and $t$ is the squared four-momentum transfer
between the incident proton and the final state neutron; 
$f_{\pi^+/p}(x_L,t)$ represents the pion flux associated with the 
splitting of a proton into a $\pi^+ n$ system 
and $\der\sigma(e\pi^+\rightarrow e'X)$ is the cross section of
the positron-pion  interaction.
 There are several parameterisations of the pion flux
 \cite{Bishari:1972tx,Holtmann:1994rs,Kopeliovich:1996iw,Przybycien:1996zb,Szczurek:1997cw} and the one used here is taken from 
\cite{Holtmann:1994rs}. 
The details of the pion flux parameterisation
are described in \cite{Aaron:2010ab}. 
Using other parameterisations of the pion flux
affects mainly the absolute normalisation by up to $30~\%$.

As was shown in \cite{Aaron:2010ab},  the best description of the 
forward neutron data is  achieved 
by a combination of events with neutrons originating from pion exchange, 
as simulated by RAPGAP-$\pi$, and events with neutrons from 
proton remnant fragmentation, simulated by DJANGOH. 
RAPGAP-$\pi$ mainly contributes at high neutron 
energies, while DJANGOH is significant at low energies.
In \cite{Aaron:2010ab} the contributions of 
 RAPGAP-$\pi$ and CDM were added using weighting factors
$0.65$ and $1.2$ for the respective predictions.
In the present analysis the best description of the neutron energy
distribution is obtained by the combination of RAPGAP-$\pi$ and CDM
using the respective weights $0.6$ and $1.4$,
or by the combination of RAPGAP-$\pi$ and LEPTO
using the respective weights $0.6$ and $0.7$.
The difference between the weighting factors for the 
combination of RAPGAP-$\pi$ and CDM in this analysis and 
in~\cite{Aaron:2010ab} is 
due to the different neutron energy range 
and the resulting different neutron energy dependence
in the two analyses.
Compared to  \cite{Aaron:2010ab} the current analysis is extended 
to much lower neutron energies, at which the contribution from the
fragmentation model dominates.

The measurements are also compared to predictions of several hadronic
interaction models which are commonly used for the simulation of cosmic ray
air shower
cascades: EPOS~LHC~\cite{Werner:2005jf,Pierog:2013ria}, 
QGSJET~01~\cite{Kalmykov:1993qe,Kalmykov:1997te},
QGSJET~II-04~
\cite{Ostapchenko:2005nj,Ostapchenko:2010vb}  
and SIBYLL~2.1\cite{Engel:1992vf,Ahn:2009wx}.
These models are based on Regge theory \cite{Collins:1977jy}, 
on the Reggeon calculus of Gribov \cite{Gribov:1968fc}
and on perturbative QCD. 
They use an unitarisation procedure to reconstruct amplitudes for
exclusive processes and to determine the total and elastic cross
sections.
Central elements of these models are the production of mini-jets and 
the formation of colour strings that fragment into hadrons.
Whereas the Regge-Gribov approximation is applied to hadrons as 
interacting objects in the case of QGSJET and SIBYLL, it is extended 
to include partonic constituents in EPOS~LHC. 
Compared to the earlier EPOS simulation~\cite{Werner:2005jf}, 
which was used for comparison with the 
previous H1 forward photon analysis \cite{Aaron:2011pe}, the new 
EPOS~LHC model~\cite{Pierog:2013ria} 
includes a  modified treatment of central diffraction and 
the diffractive remnant in order to reproduce rapidity gap 
measurements at the LHC.
The CR models also differ
in the treatment of saturation effects at high parton densities at small 
Bjorken-$x$
and in the treatment of the hadronic remnants in collisions.
 The programs are interfaced with the PHOJET program~\cite{Engel:1995yda}
 for the generation of the $ep$ scattering kinematics.
It was pointed out \cite{ostapchenko1} that the hadronic interaction models
have been developed for hadron-hadron interactions and therefore
the simulation of DIS events might
be affected by the superfluous contribution of 
multi-parton interactions. In order to investigate this assumption,
the QGSJET~01 model has been modified \cite{ostapchenko2} to exclude 
the multi-parton interactions. In the comparison with the  measurements
this modified model is denoted as `QGSJET~01~(no~mi)'.

The measured distributions may contain background arising from several
sources.  The background from photoproduction processes
is estimated using the PHOJET MC generator. It is found to
be about $1\%$ on average and is subtracted from the data distributions
bin-by-bin.
Background from misidentification of photons or
neutrons in the FNC is estimated
 from the DJANGOH ~MC simulation 
 to be $2\%$ on average
 and is subtracted from the data distributions bin-by-bin.
Background also arises from a random
coincidence of DIS events, causing activity in the central detector, 
with a beam-related background signal in the FNC, produced
from the interaction of another beam proton with a positron or
with residual gas in the beampipe. 
This contribution is estimated by combining DIS events with FNC clusters
originating from interactions in the bunch-crossings adjacent
to the bunch-crossings of the DIS events. 
It is found to be smaller than $1\%$ and is neglected.
%

The MC simulations are 
used to correct the distributions 
at the level of reconstructed particles
back to the hadron level on a bin-by-bin basis.
The correction factors are determined from  the 
MC simulations as the ratios of the normalised cross sections 
obtained from particles at hadron level without QED radiation
to the normalised cross sections calculated using reconstructed particles
and including QED radiation effects.
For the forward photon analysis 
the average of the correction factors determined from LEPTO and CDM 
is used. 
For the forward neutron analysis the correction factors are calculated
using the  combination of RAPGAP-$\pi$ and CDM  simulations,
with the weighting factors $0.6$ and $1.4$, as described above. 
The size of the 
correction  factors varies between $2$ and $3.5$ for the forward photon
and between $2$ and $6$ for the forward neutron $x_F$ distributions, 
and is about $3.2$ for the $W$  distribution in both cases.
The correction factors are large mainly  due to
 the non-uniform azimuthal acceptance of the FNC, 
which is about $30\%$ on average.
The bin purity, defined as the fraction
of events reconstructed in a particular bin that originate from
the same bin on hadron level,  varies between $75\%$ and $95\%$.

\subsection{Systematic uncertainties}

Several sources of experimental uncertainties are considered and
their effect on the measured cross section is quantified. The systematic
uncertainties on the cross section measurements are determined using
MC simulations, by propagating the corresponding
uncertainty through the full analysis chain.

As the cross sections are normalised to the inclusive DIS cross
section measured in this analysis, some important systematic
uncertainties,
 such as those involving the trigger efficiency and the integrated luminosity
 and those related to the reconstruction of the scattered positron
 and the hadronic final state 
are largely reduced or cancel.
The following sources are considered for both the DIS sample and the 
FNC samples:
\begin{itemize}
\item
The uncertainty on the measurements of the scattered positron energy ($1\%$).
\item
The uncertainty on the measurements of the scattered positron angle ($1$~mrad).
\item
The uncertainty on the measurements of the energy of the hadronic final state ($4\%$). 
\item 
The uncertainty on the trigger efficiency ($1\%$).
\end{itemize}
These uncertainties are strongly correlated between the DIS and the 
forward photon and neutron samples. 
The resulting combined uncertainty of the cross section is 
about $2\%$ on average and
is considered as uncorrelated between the measurement points.

Several sources of uncertainties related to the reconstruction of the 
forward photons and neutrons in the FNC are considered:
\begin{itemize}
\item
The acceptance of the FNC calorimeter is
defined by the interaction point and the geometry of the HERA magnets
and is determined using MC simulations.
The uncertainty of the impact position of the particle on the FNC,
due to beam inclination and the uncertainty on the FNC position,
is estimated to be $5$~mm.  This results in uncertainties on the 
FNC acceptance determination of up to $15\%$ for the $x_F$ distributions. 
\item
The absolute electromagnetic energy scale of the FNC is known to a
precision of $5\%$~\cite{Aaron:2011pe}.
 This leads to an uncertainty of $1\%$ on the forward photon
cross section measurement at low energies, increasing to $11\%$ for the
largest $x_F$ values. 
\item
The uncertainty in the neutron detection efficiency and the $2\%$ uncertainty 
on the absolute hadronic energy scale of the FNC~\cite{Aaron:2010ab} 
lead to 
systematic errors 
on the cross section of  $2\%$ and up to $10\%$, respectively.
\end{itemize}
These systematic uncertainties related to the FNC
are strongly correlated between measurement bins and
mainly contribute to the overall normalisation uncertainty.
For the normalised cross sections studied as a function of $W$
all above-mentioned FNC related systematic uncertainties 
contribute to a normalisation uncertainty of approximately $7\%$.

In the procedure of correcting the measurements to the hadron level, using
MC simulations, the following sources of systematic uncertainties are 
considered:
\begin{itemize}
\item
The systematic uncertainty arising from the radiative corrections 
and the model dependence of the data correction for the forward neutron cross 
section is estimated by varying
the DJANGOH and RAPGAP-$\pi$ scaling factors 
within values permitted by the data and by switching between
the CDM and LEPTO  models within DJANGOH.
 The resulting uncertainty on the cross section is typically $4\%$,
increasing to $5\%$ at lowest and highest $x_F$ values.
\item
For the forward photon cross section the systematic uncertainty,
taken as the difference of the acceptance corrections calculated
using the LEPTO and CDM models, increases from $1\%$ at low $x_F$
to $7\%$ at higher $x_F$.
\item
The use of different parton
distribution functions in the MC simulation results in a negligible
change of the correction factors.
\item
A $2\%$ uncertainty is attributed to the bin-by-bin subtraction
of background 
arising from the wrong identification of photon
and neutron candidates in the FNC and from photoproduction
processes.
\end{itemize}
These uncertainties
are assumed to be equally
shared between correlated and uncorrelated parts. 

The systematic uncertainties shown in the figures and tables 
are calculated using the quadratic sum of all contributions, 
which may vary from point to point. 
They are significantly larger than the statistical uncertainties 
in all measurement bins.
The total systematic uncertainty for the normalised cross section
measurements ranges between $8\%$ and $22\%$ for the $x_F$ dependent cross
sections and is about $8\%$ for the $W$ dependent cross sections.

\section{Results}

\subsection{Normalised cross sections as a function of $W$}
\label{sec:resultW}

\begin{figure}[h]
\epsfig{file=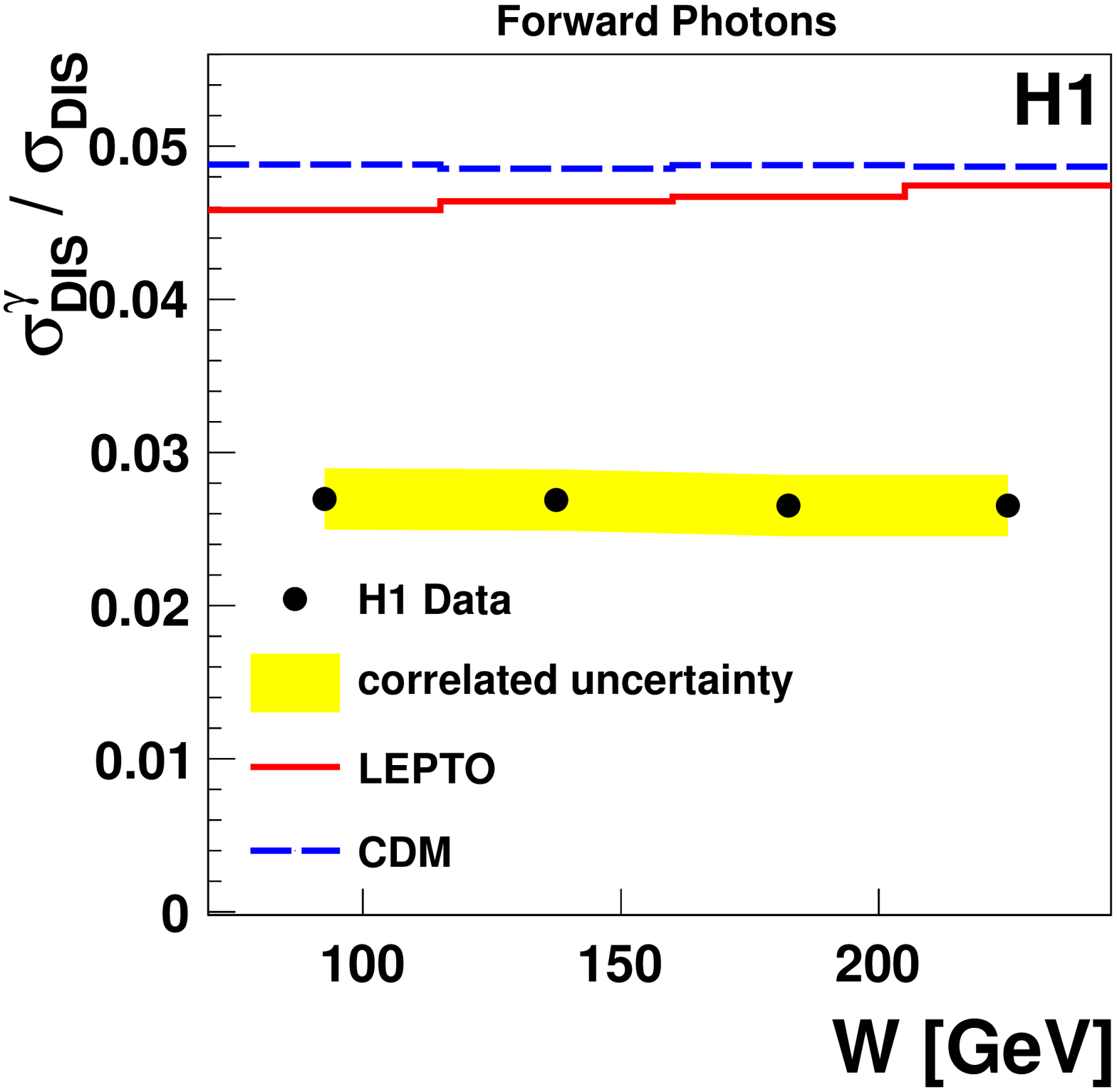,width=79mm}
\epsfig{file=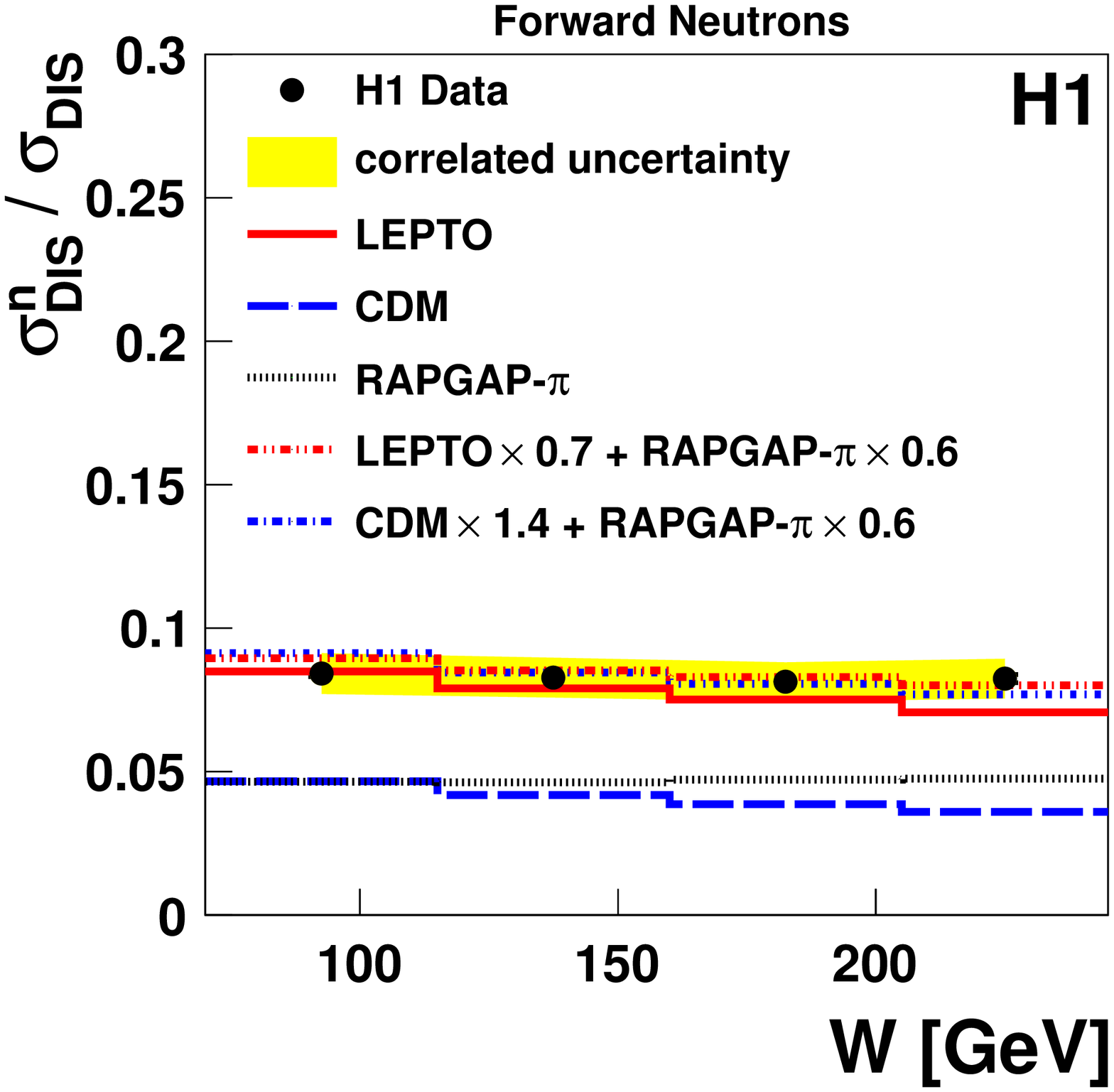,width=79mm}
\caption{
The fraction of DIS events with forward photons {\bf (a)} and
forward neutrons {\bf (b)} as a function of $W$
in the kinematic region given in Table~\ref{tab:selection}.
Also shown are the predictions of the LEPTO (solid line) and CDM (dashed line)
MC models. In the case of forward neutron production, the predictions of
RAPGAP-$\pi$ and the linear combinations of LEPTO and RAPGAP-$\pi$, as 
 well as  CDM and RAPGAP-$\pi$ are also shown.
}

\vspace*{-92mm}
{
\bf \large
\hspace*{67mm}(a) 
\hspace*{74mm}(b) 
		
}
\label{figW}
\end{figure}

The ratios of the forward photon and forward neutron production
cross sections to the inclusive DIS cross section, 
 $ \sigma^{\gamma,n}_{DIS}/\sigma_{DIS}$, 
in the kinematic range
$6<Q^2<100~\GeVS$ and $0.05<y<0.6$ and as a function of $W$
are presented in Tables~\ref{tab:table1} and \ref{tab:table2}
and are shown in  Figure~\ref{figW}.
Within uncertainties the measured ratios
 are consistent  with constant values of  about
$0.027$ (forwards photons) and $0.083$ (forward neutrons).
In other words, within uncertainties the $W$ dependence
   of the cross section is independent of the presence of a forward
   neutron or a forward photon, as predicted by the limiting
   fragmentation hypothesis ~\cite{Benecke:1969sh,Chou:1994dh}.

In Figure~\ref{figW} the MC model calculations are compared to the measurements.  
Both CDM and LEPTO predict 
a forward photon rate of about $70\%$ higher than observed.
A similar excess was observed earlier \cite{Aaron:2011pe}.
The photon production rate as a function of $W$ is rather flat in CDM
and shows a slight increase with $W$ in LEPTO. 
The shape of the $W$ distribution is in both models consistent with the data,
within errors.
%

The rate of forward neutron production predicted by LEPTO is consistent
with the data, while  CDM predicts a much lower rate. However, as was 
shown in the previous measurement~\cite{Aaron:2010ab}, 
the energy distribution
of forward neutrons can be described by MC simulation only if this
includes contributions both from standard fragmentation as simulated
in DJANGOH, and from a pion exchange mechanism as explicitly simulated in
RAPGAP-$\pi$ but not included in DJANGOH. 
In  Figure~\ref{figW}b the combinations of the RAPGAP-$\pi$ and DJANGOH
simulations, as described in section \ref{sec:mc},  
are compared to the measurement.
The weighting factors $1.4$ for the CDM, $0.7$ for the LEPTO and $0.6$ for the
RAPGAP-$\pi$
predictions are determined by fitting the observed neutron
energy distributions integrated over the full $W$ range.
The cross sections for inclusive DIS events, used for the normalisation 
of the forward neutron cross sections, $\sigma_{DIS}$, are taken from
the CDM and LEPTO simulations without additional weights. 
The model combination describes the observed $W$ dependence well.
It is remarkable that 
the factors for
the CDM and LEPTO contributions differ by a factor two 
($1.4$ and $0.7$, respectively).
It is also notable that the CDM model, which overestimates
the rate of forward photons by about $70\%$, has to be scaled up 
in the combination to describe the forward neutron data.

\begin{figure}[h]
\epsfig{file=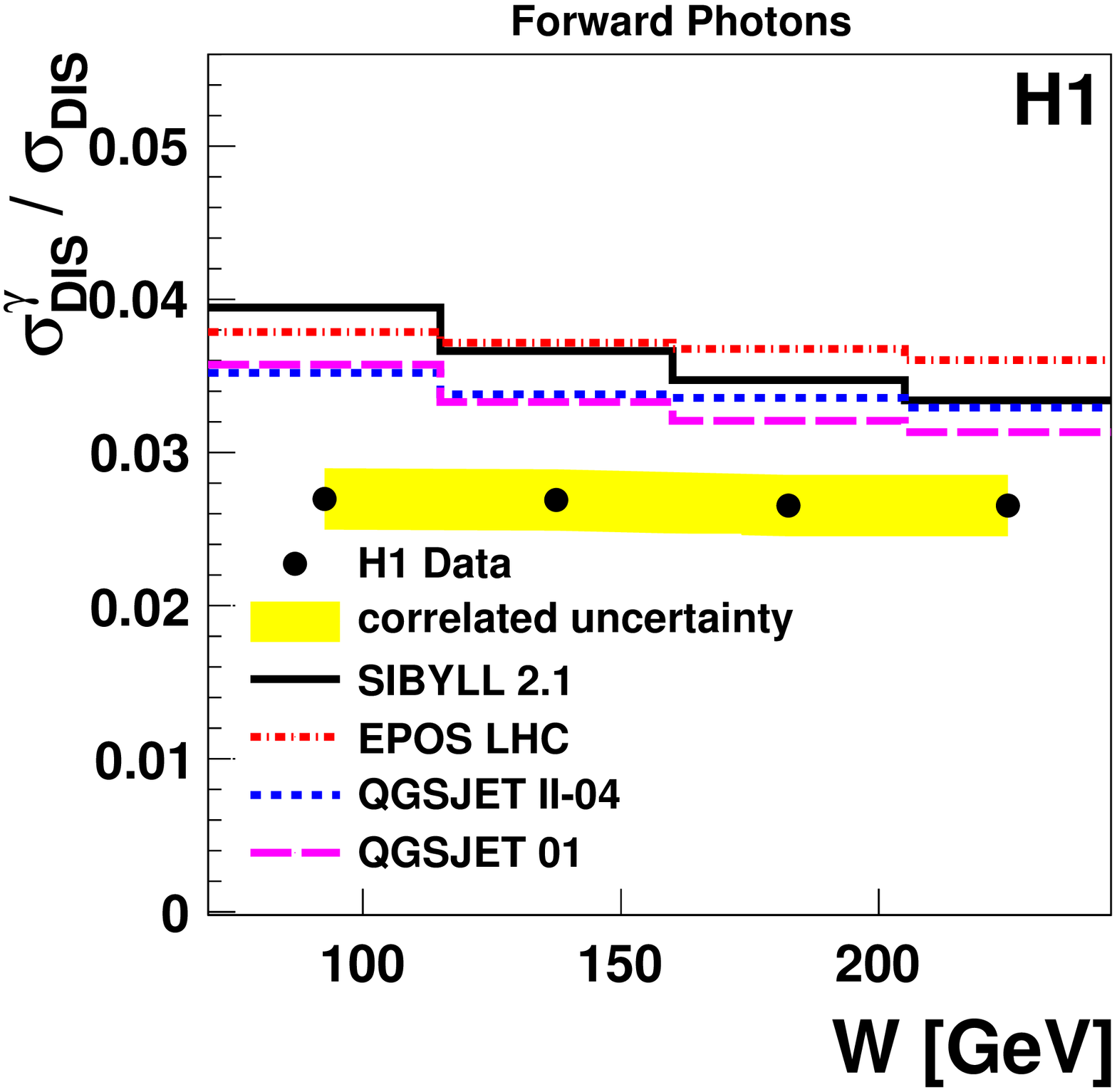,width=79mm}
\epsfig{file=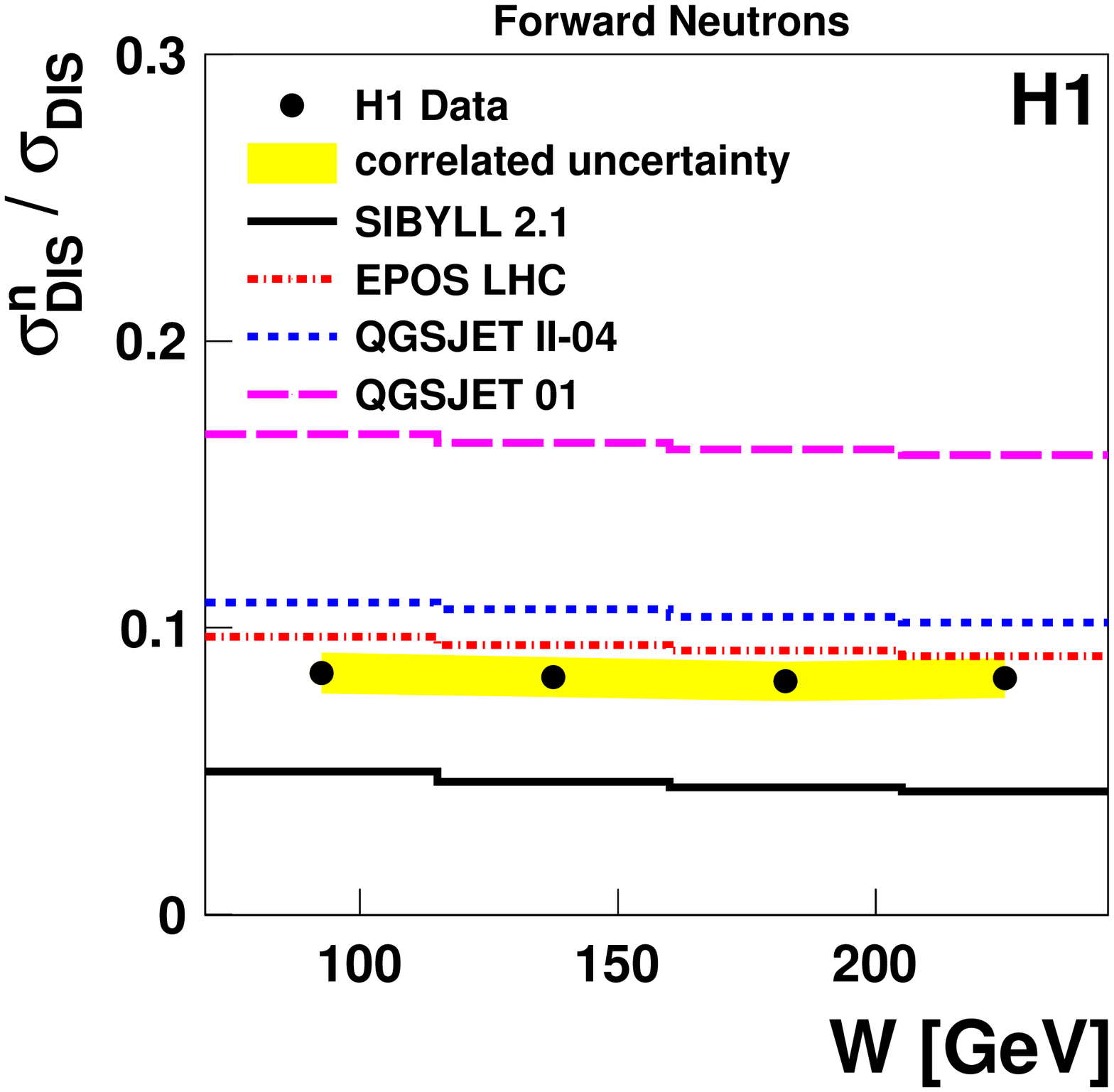,width=79mm}
\caption{
The fraction of DIS events with forward photons {\bf (a)} and
forward neutrons {\bf (b)} as a function of $W$
in the kinematic region given in Table~\ref{tab:selection}.
Also shown are the predictions of the cosmic ray hadronic interaction
models SIBYLL~2.1 (solid line), QGSJET~01 (dashed line), QGSJET~II-04 (dotted line)
and EPOS~LHC (dash-dotted line).
}
\vspace*{-85mm}
{
\bf \large
\hspace*{67mm}(a) 
\hspace*{74mm}(b) 
}
\label{figWcos}
\end{figure}

In  Figure~\ref{figWcos} 
predictions of various cosmic ray hadronic interaction models (EPOS~LHC, 
SIBYLL~2.1 and the two versions of QGSJET) are compared to the measured
normalised cross sections as a function of $W$.
The CR model predictions show significant differences in
absolute values,
for both forward photons and forward neutrons. 
For photons all models predict too high rates by
$30$ to $40\%$, and these rates, with the exception of EPOS~LHC, show a slight
decrease with increasing $W$, not confirmed by  data.
For forward neutrons all CR predictions show a $W$ independent behaviour,
in accordance with the measured $W$ dependence.
The QGSJET~01 model predicts a much too high and SIBYLL~2.1 a much
too low neutron rate,
while the EPOS~LHC and QGSJET~II-04 models are closer to the measurement.


\subsection{Normalised cross sections as a function of \boldmath $x_F$ and test of Feynman scaling}
\begin{figure}[h]
\epsfig{file=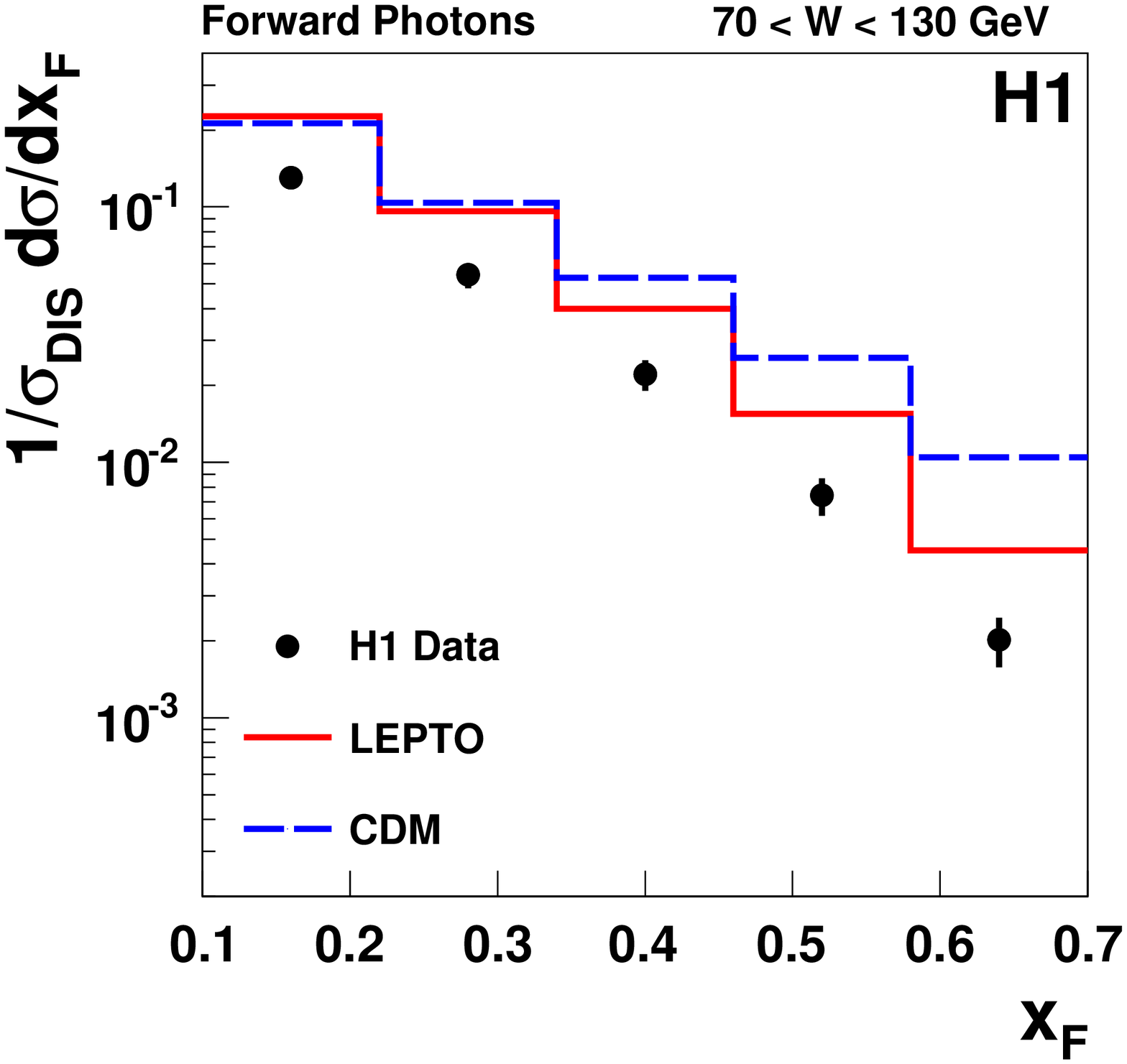,width=67mm}
\epsfig{file=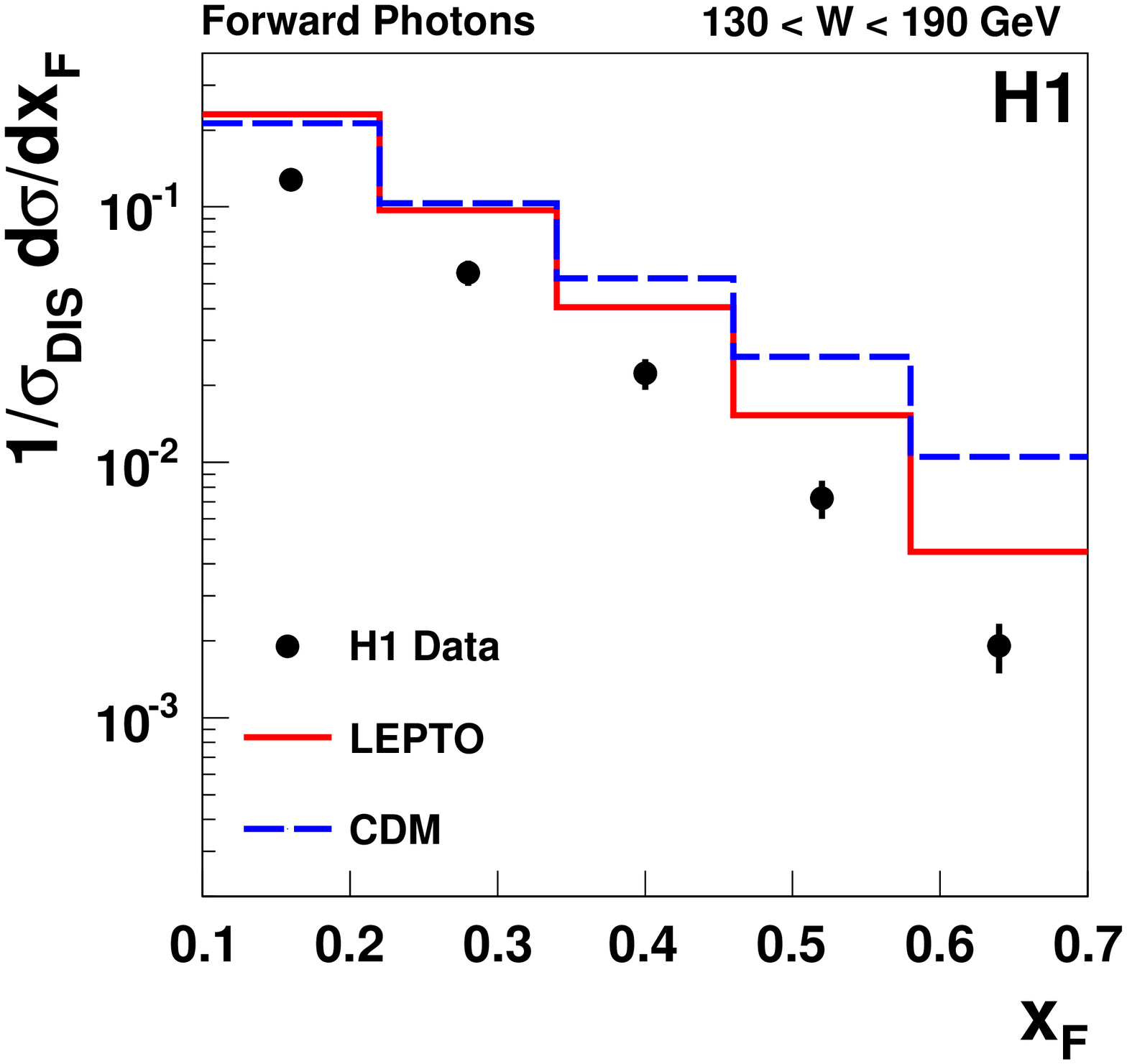,width=67mm}
\epsfig{file=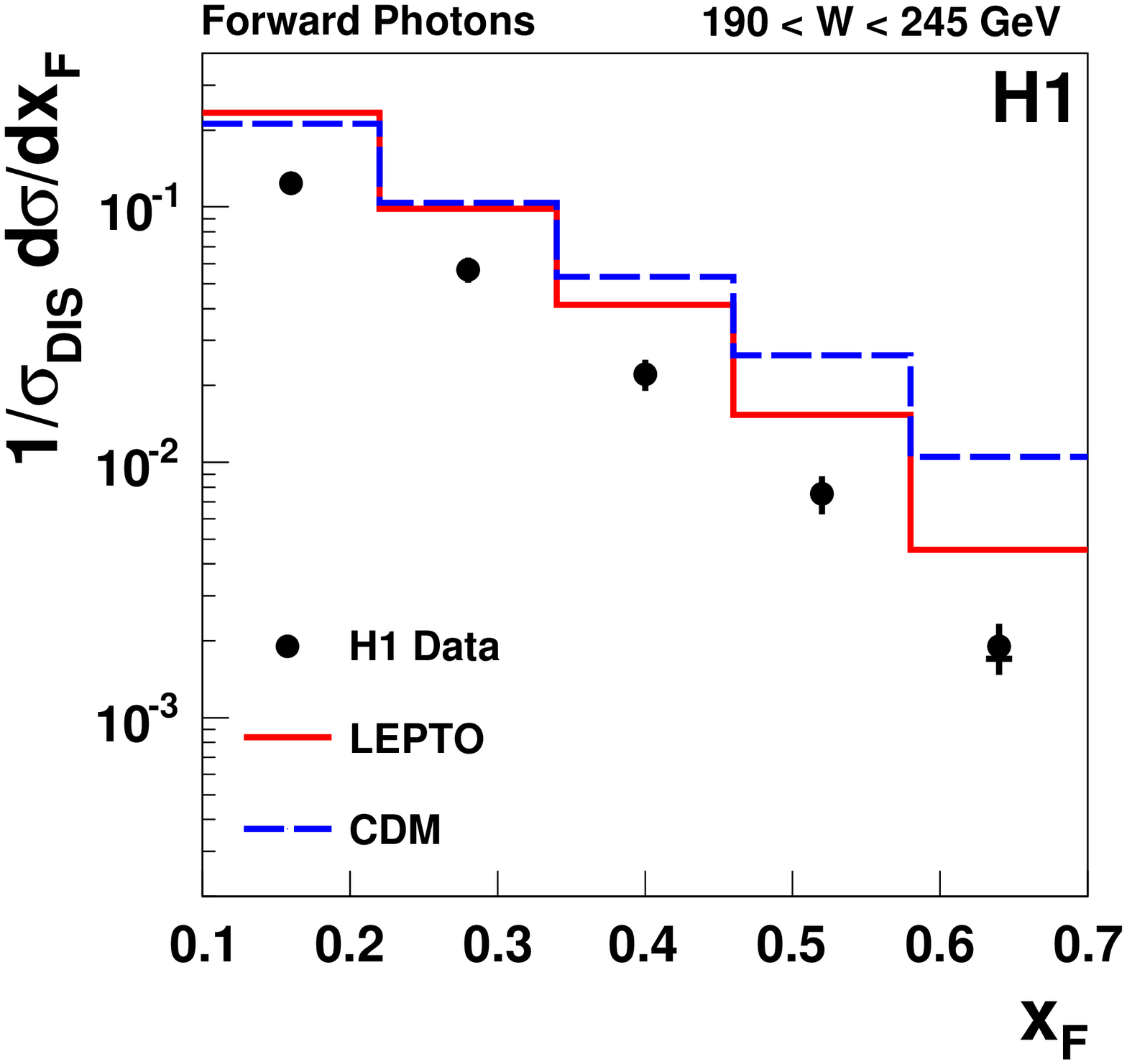,width=67mm}
\caption{
Normalised cross sections of forward photon production in DIS
as a function of $x_F$ in three $W$ intervals
in the kinematic region given in Table~\ref{tab:selection}.
The error bars show the total experimental uncertainty, calculated using the 
quadratic sum of the statistical and systematic uncertainties.
Also shown are the predictions of the LEPTO (solid line) and CDM (dashed line)
MC models.
}
\label{fig:xfgmc}
\end{figure}

\begin{figure}[h]
\centering
\epsfig{file=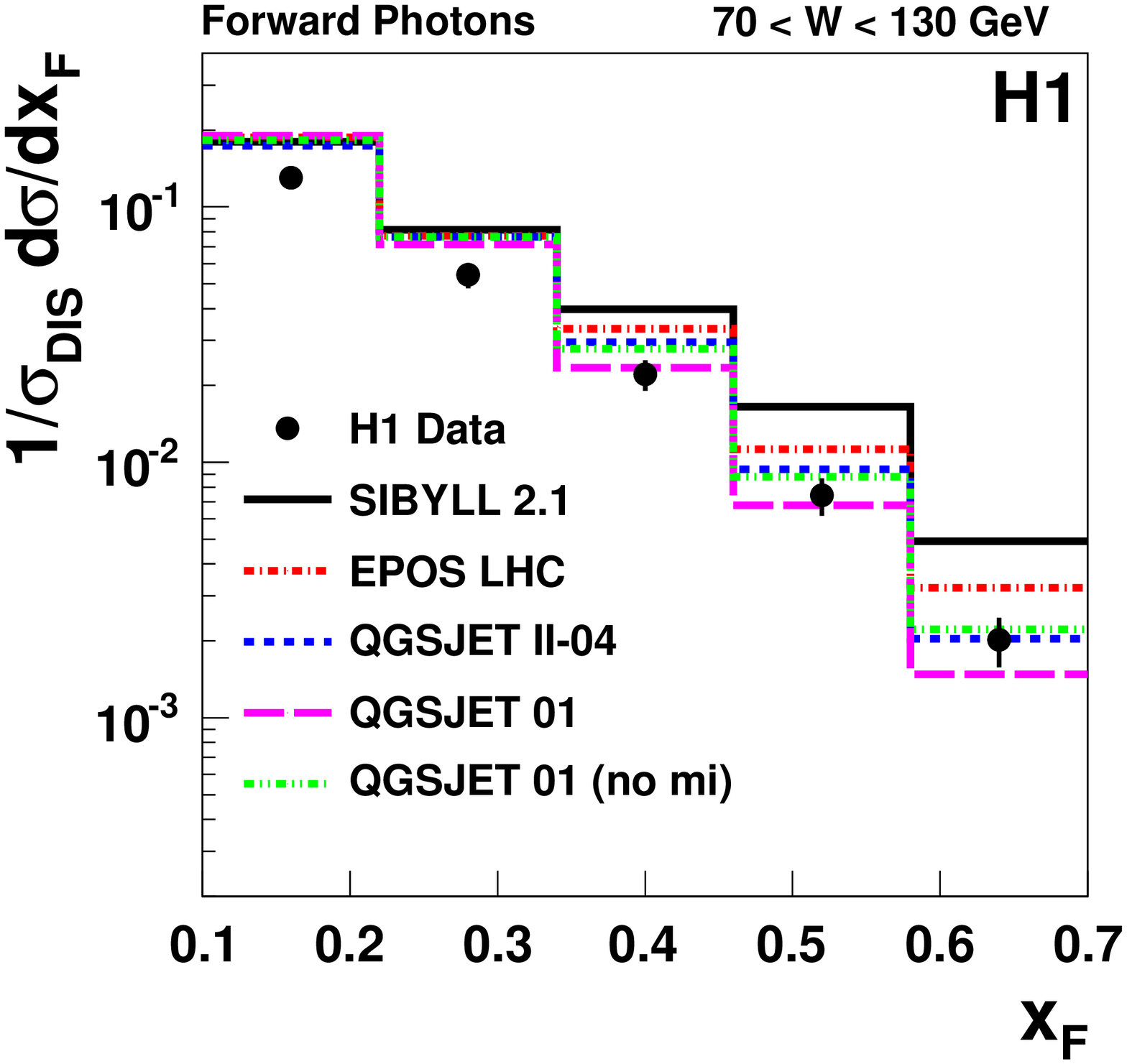,width=65mm}
\epsfig{file=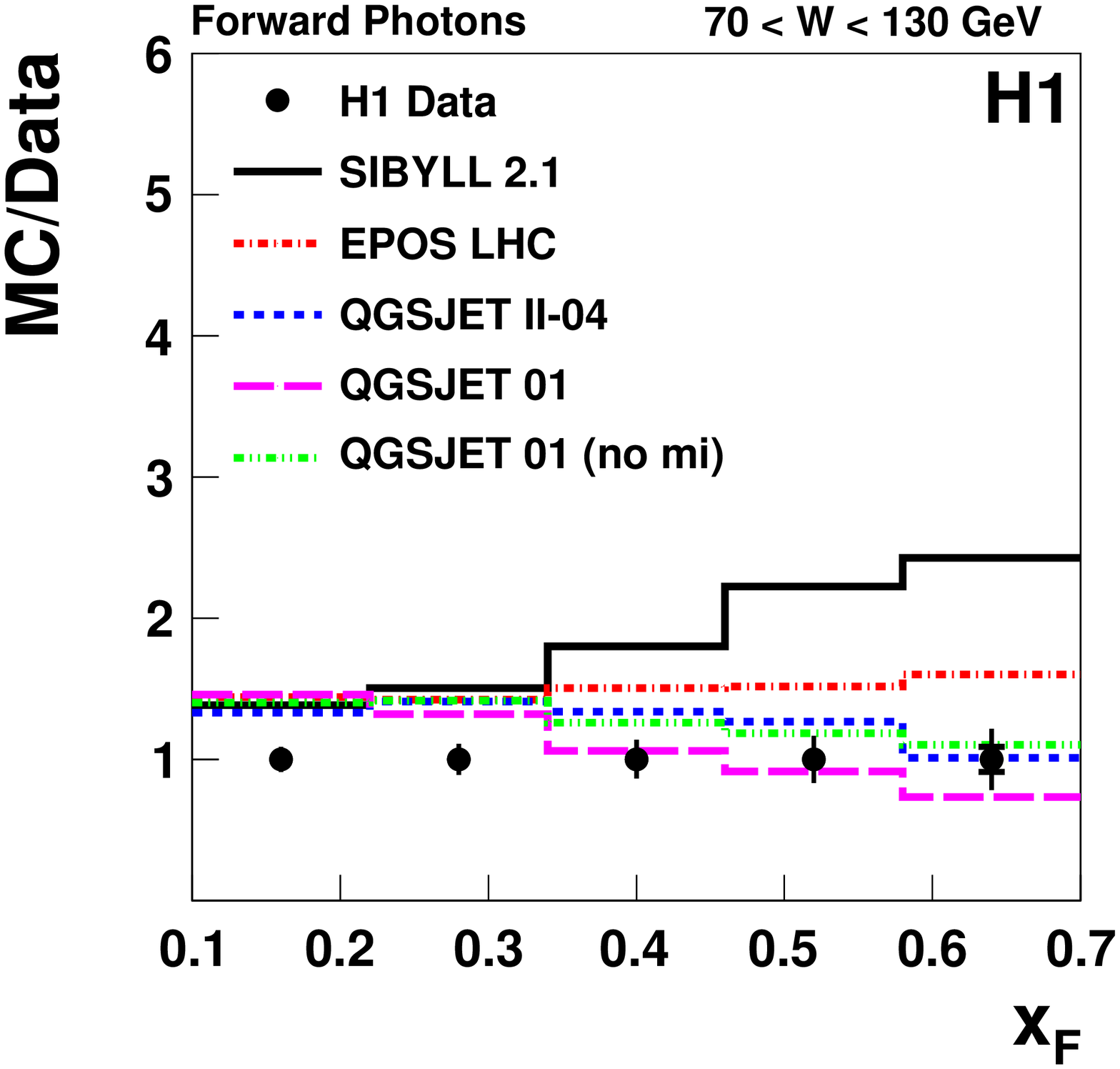,width=65mm}

\epsfig{file=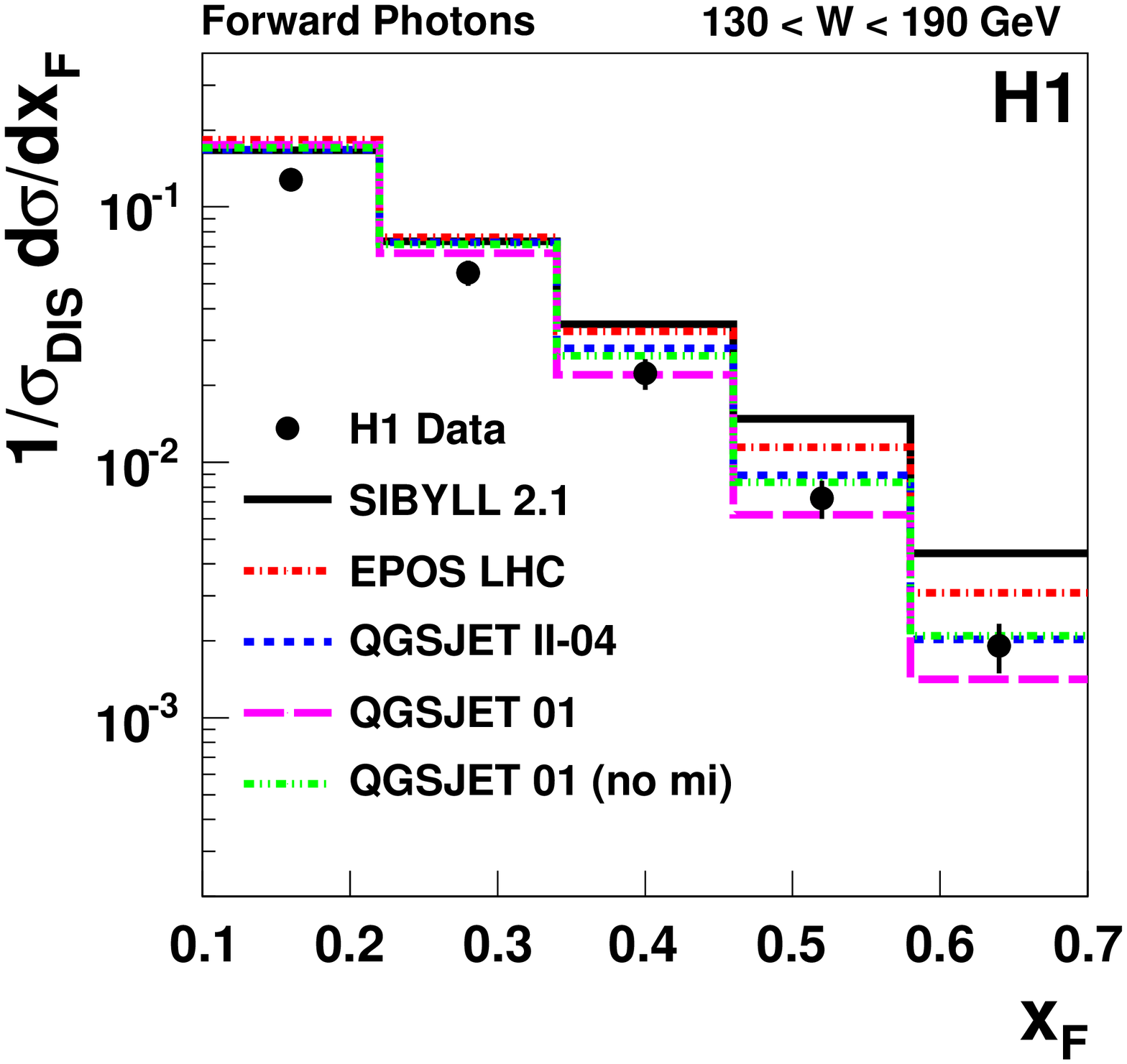,width=65mm}
\epsfig{file=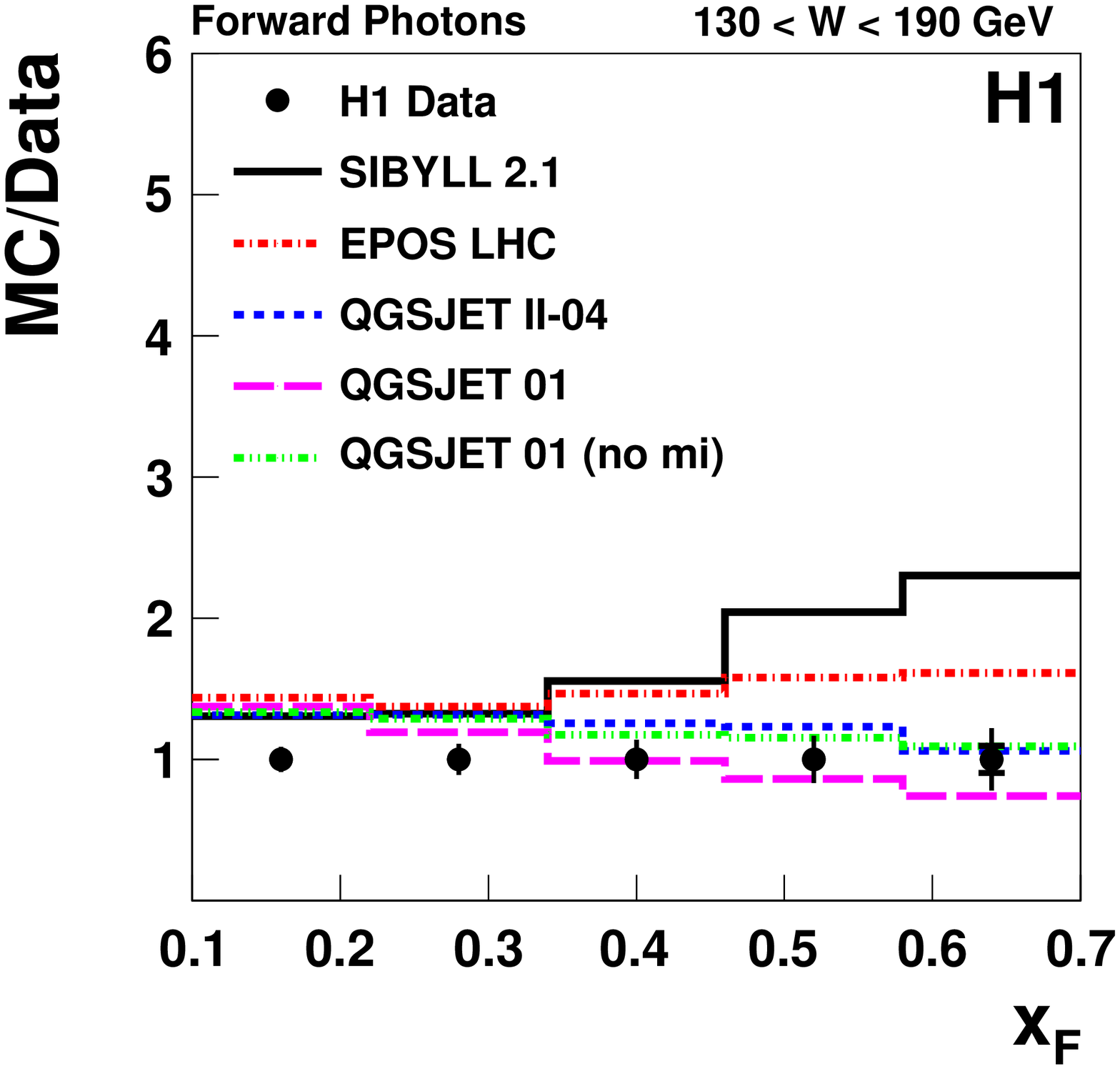,width=65mm}

\epsfig{file=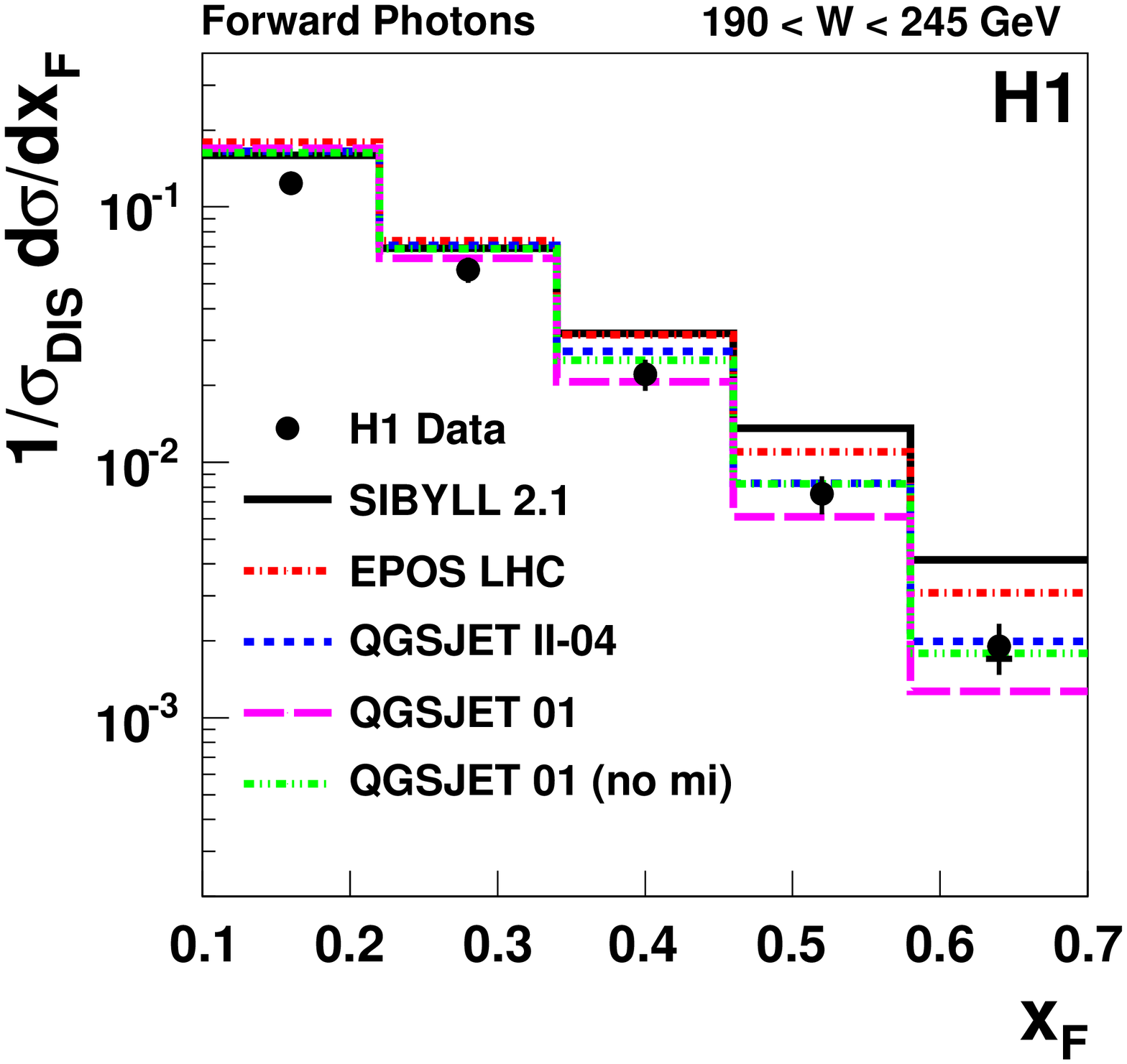,width=65mm}
\epsfig{file=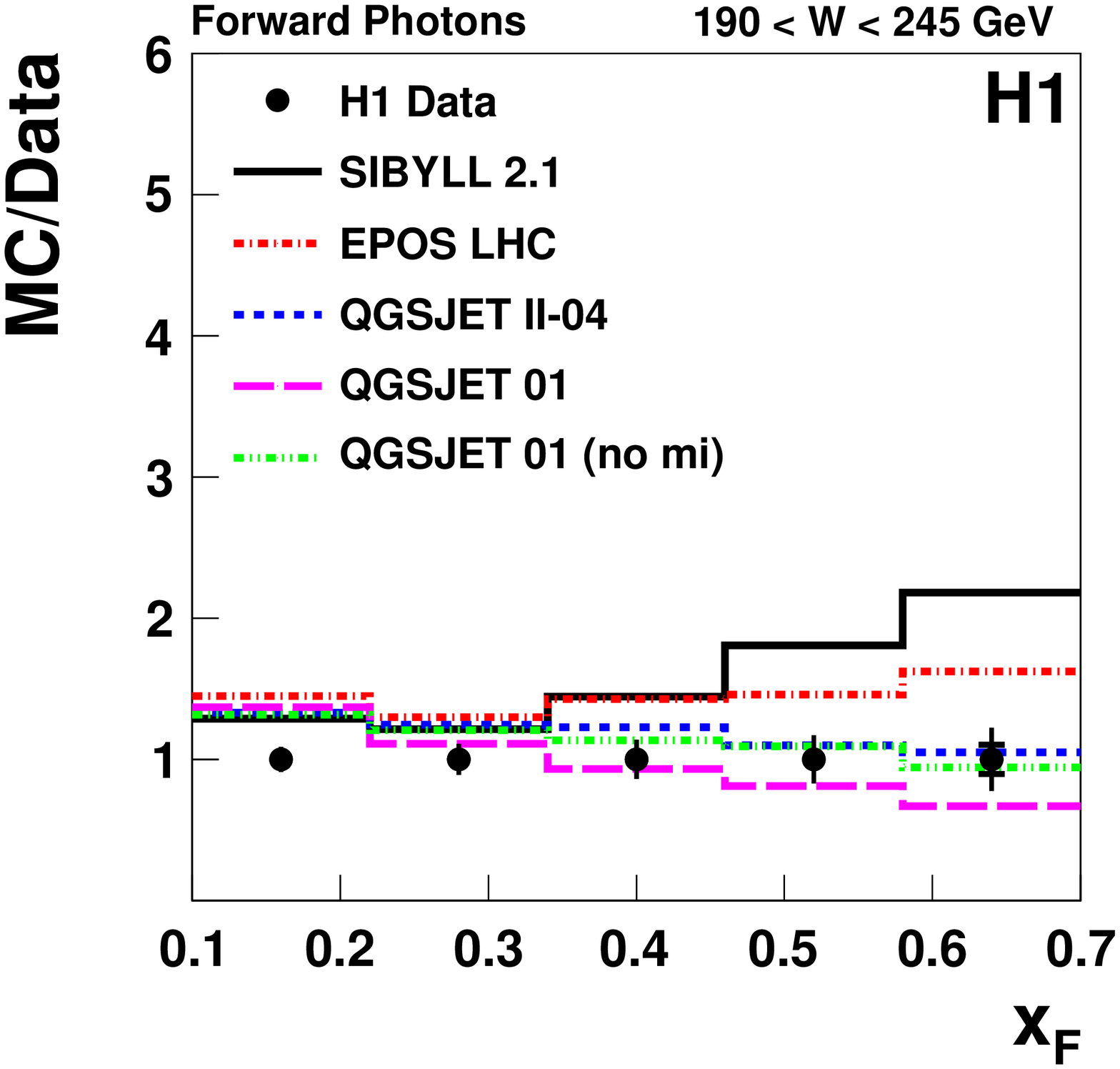,width=65mm}

\caption{
Normalised cross sections of forward photon production in DIS
as a function of $x_F$ in three $W$ intervals
in the kinematic region given in Table~\ref{tab:selection}.
The inner error bars show the statistical uncertainty, while
the outer error bars show the total experimental uncertainty, 
calculated using the
quadratic sum of the statistical and systematic uncertainties.
Also shown are the predictions of the cosmic ray hadronic interaction
models SIBYLL~2.1 (solid line), QGSJET~01 (dashed line), 
QGSJET~01~(no~mi) (dash-double dotted line),  
QGSJET~II-04 (dotted line) and EPOS~LHC (dash-dotted line).
In the right column the ratios of the CR model predictions to the data are shown.
}
\label{fig:xfgcos}
\end{figure}

The measured normalised differential cross sections,
  $ {1}/\sigma_{DIS}~  {\rm d}\sigma/{\rm d}x_F$,
 of the most energetic photon  
are presented as a function of $x_F$ 
in  Table~\ref{tab:table3} and in Figures~\ref{fig:xfgmc} and \ref{fig:xfgcos}
for the kinematic region defined in Table~\ref{tab:selection}.
In order to study the energy dependence of the $x_F$ distributions,
these cross sections are measured in three $W$ intervals.

The normalised differential cross sections as a function of $x_F$ 
are similar for the three $W$ ranges.
As shown in Figure~\ref{fig:xfgmc}
and already seen in the comparison of the $W$ dependence,
the LEPTO and CDM models predict a rate of forward photons about $70\%$ 
higher than measured.
The shapes of the measured distributions are well described by
LEPTO, while the CDM description is very poor
by showing a significantly harder spectrum than observed in data.
In Figure~\ref{fig:xfgcos} the predictions of the CR hadronic 
interaction models are compared to the same measurements.
Large differences between the CR models are observed, both in shape and
in normalisation.
All models tested here overestimate the forward photon 
rate by $30\%$ to $40\%$ at low $x_F$.
The EPOS~LHC model describes the shapes of the photon $x_F$
distributions well. 
The SIBYLL~2.1 model
predicts a harder $x_F$ dependence,
while the spectra obtained from the different variants of QGSJET are
softer than observed in the data.
%
Forward photon and neutral pion measurements at the LHC also revealed 
differences with respect to a similar selection of CR 
models~\cite{Adriani:2011nf,Adriani:2012ap,Adriani:2012fz}.

\begin{figure}[h]
\epsfig{file=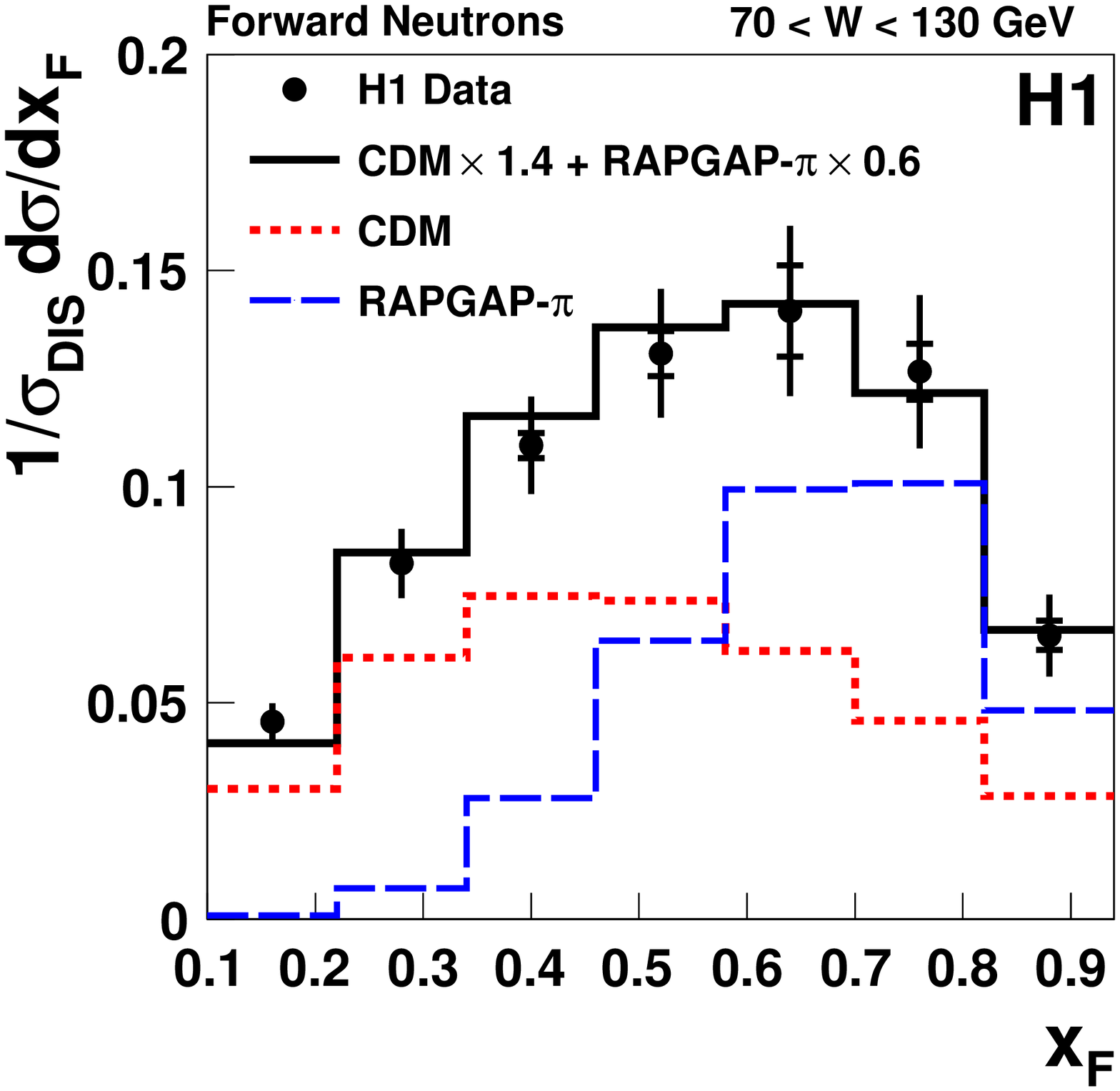,width=67mm}
\epsfig{file=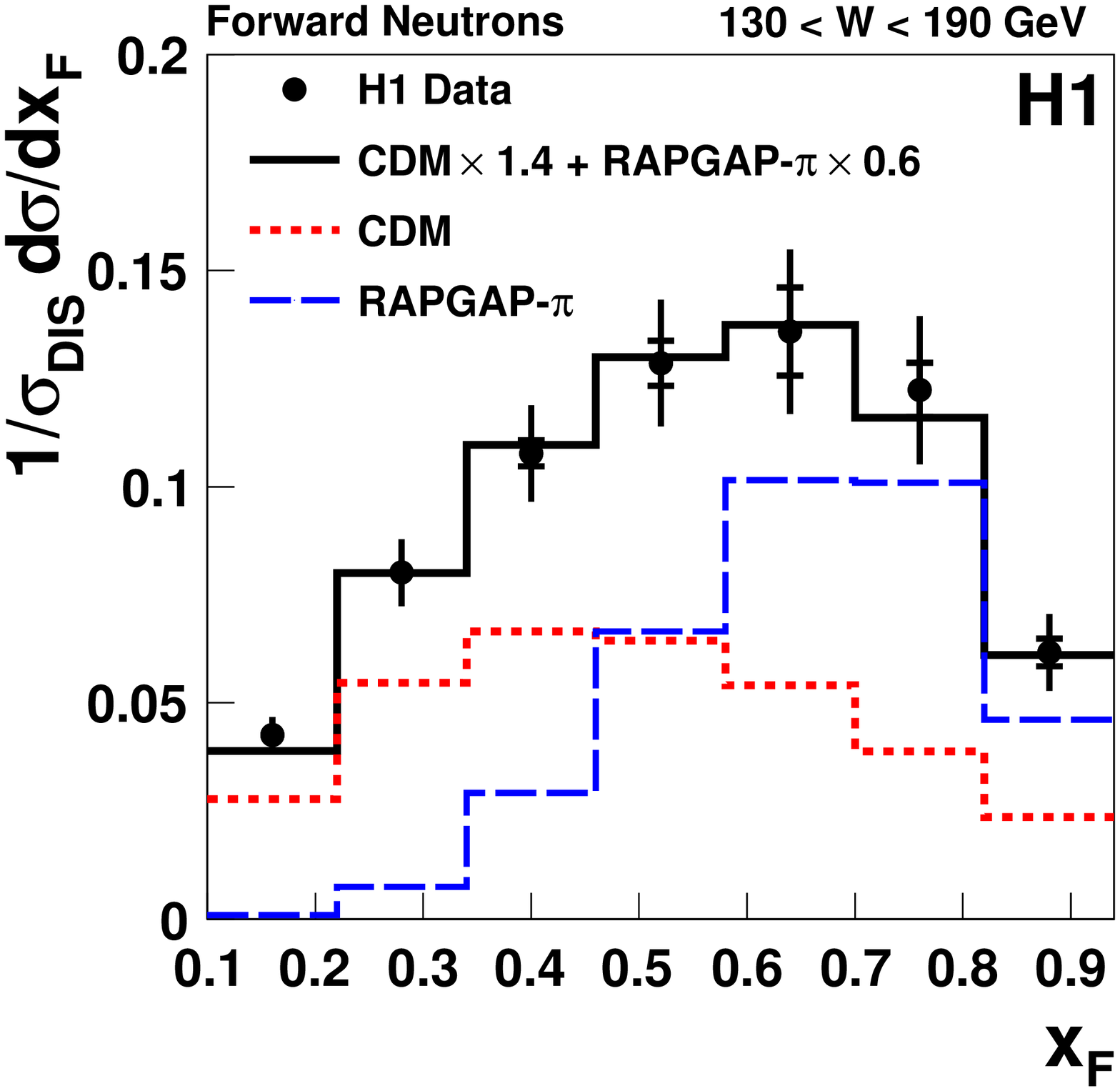,width=67mm}
\epsfig{file=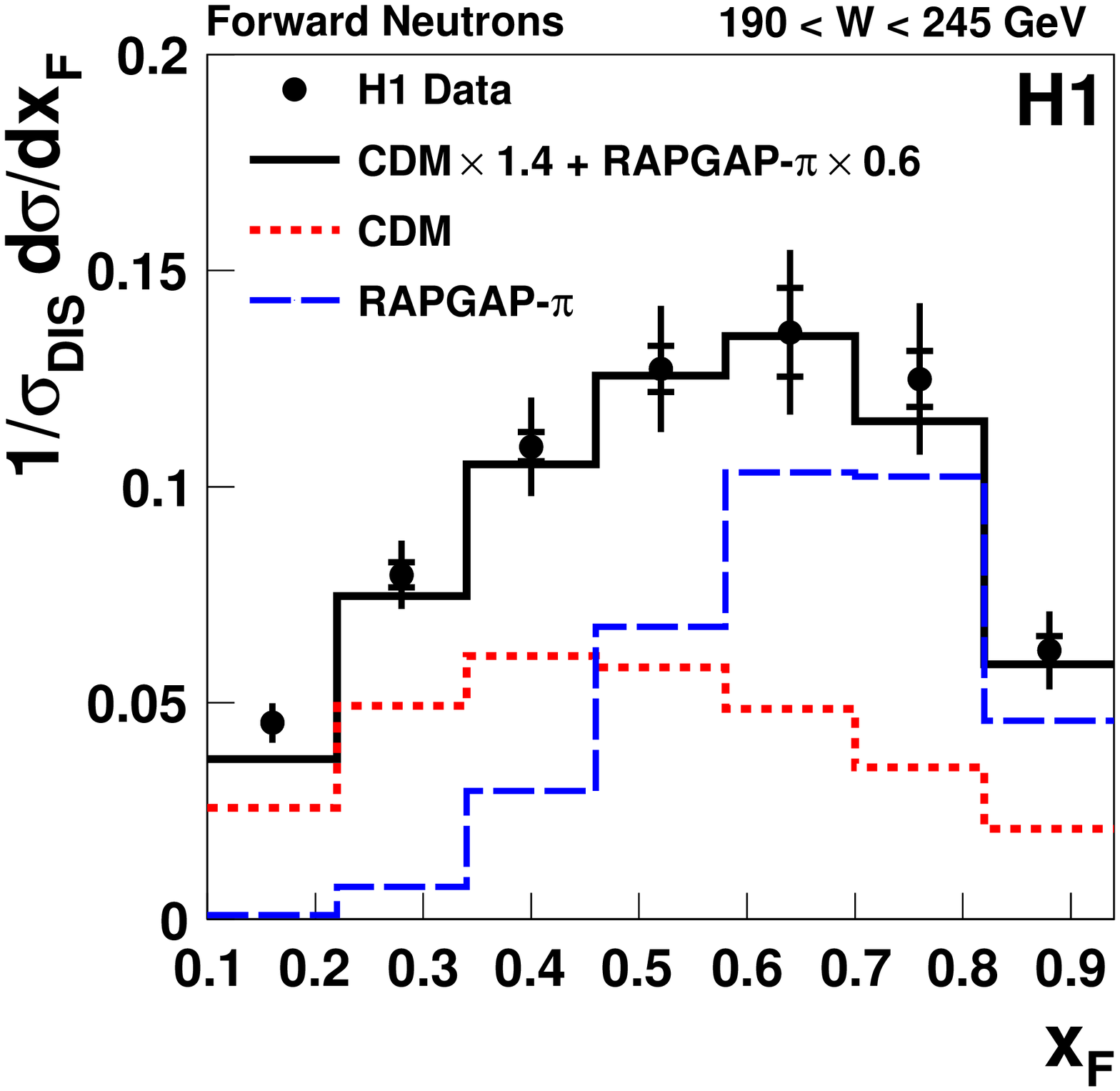,width=67mm}
\caption{
Normalised cross sections of forward neutron production in DIS
as a function of $x_F$ in three $W$ intervals
in the kinematic region given in Table~\ref{tab:selection}.
The inner error bars show the statistical uncertainty, while
the outer error bars show the total experimental uncertainty, 
calculated using the
quadratic sum of the statistical and systematic uncertainties.  
Also shown are the predictions of CDM (dotted line), RAPGAP-$\pi$ (dashed line)
and a linear combination of CDM and  RAPGAP-$\pi$ predictions (solid line).
}
\label{fig:xfn}
\end{figure}
\begin{figure}[h]
\centering
\epsfig{file=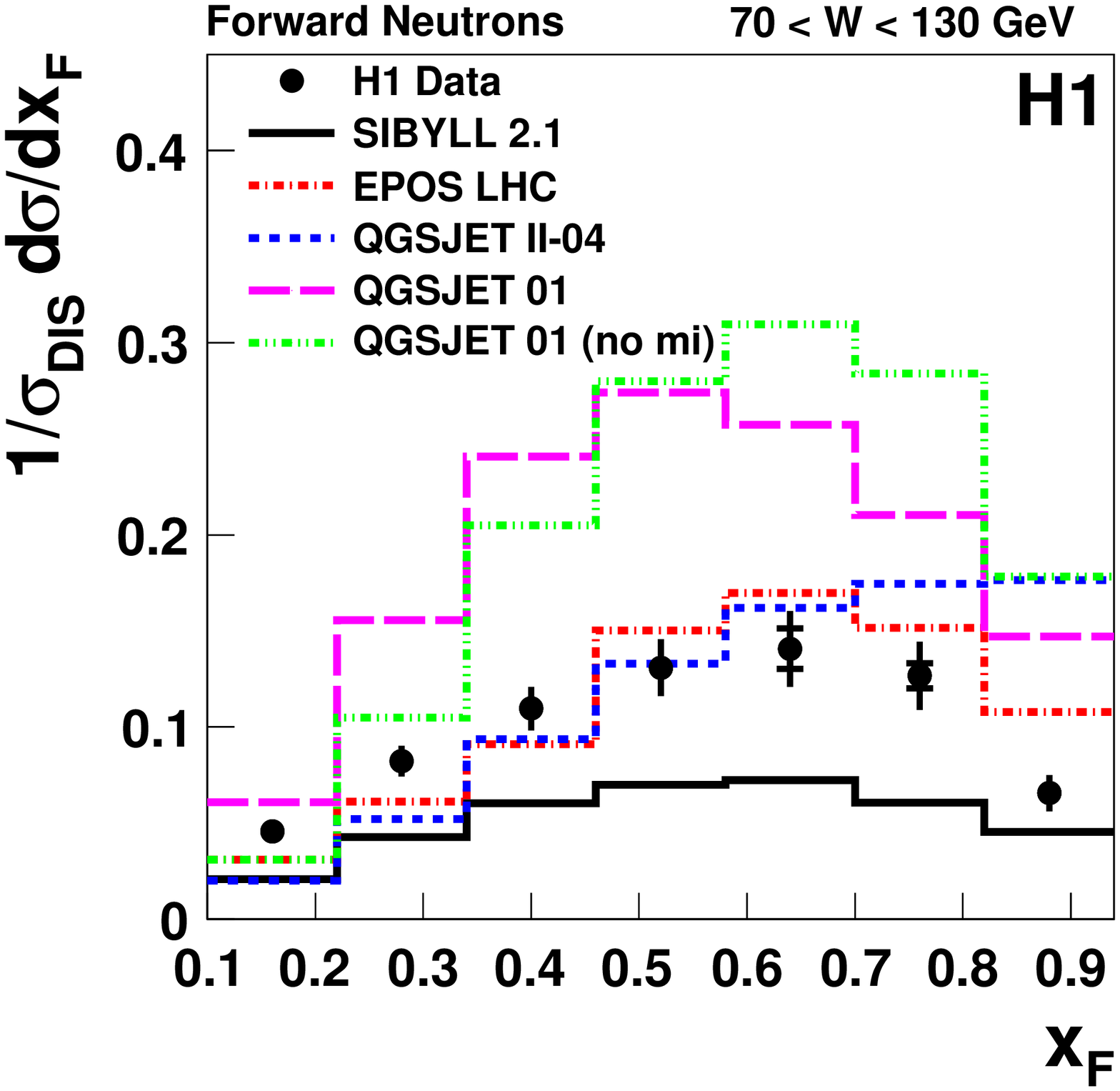,width=65mm}
\epsfig{file=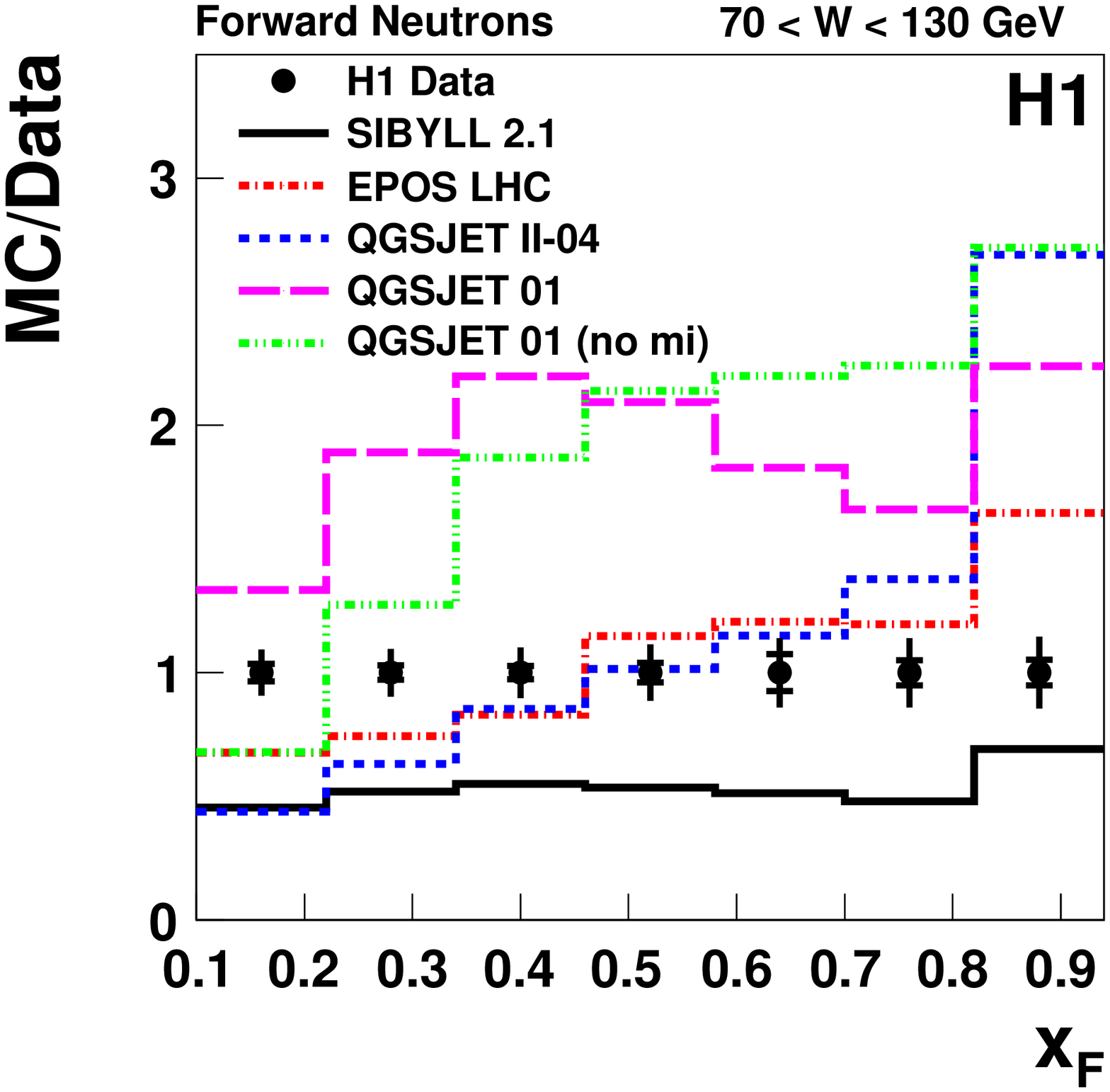,width=65mm}

\epsfig{file=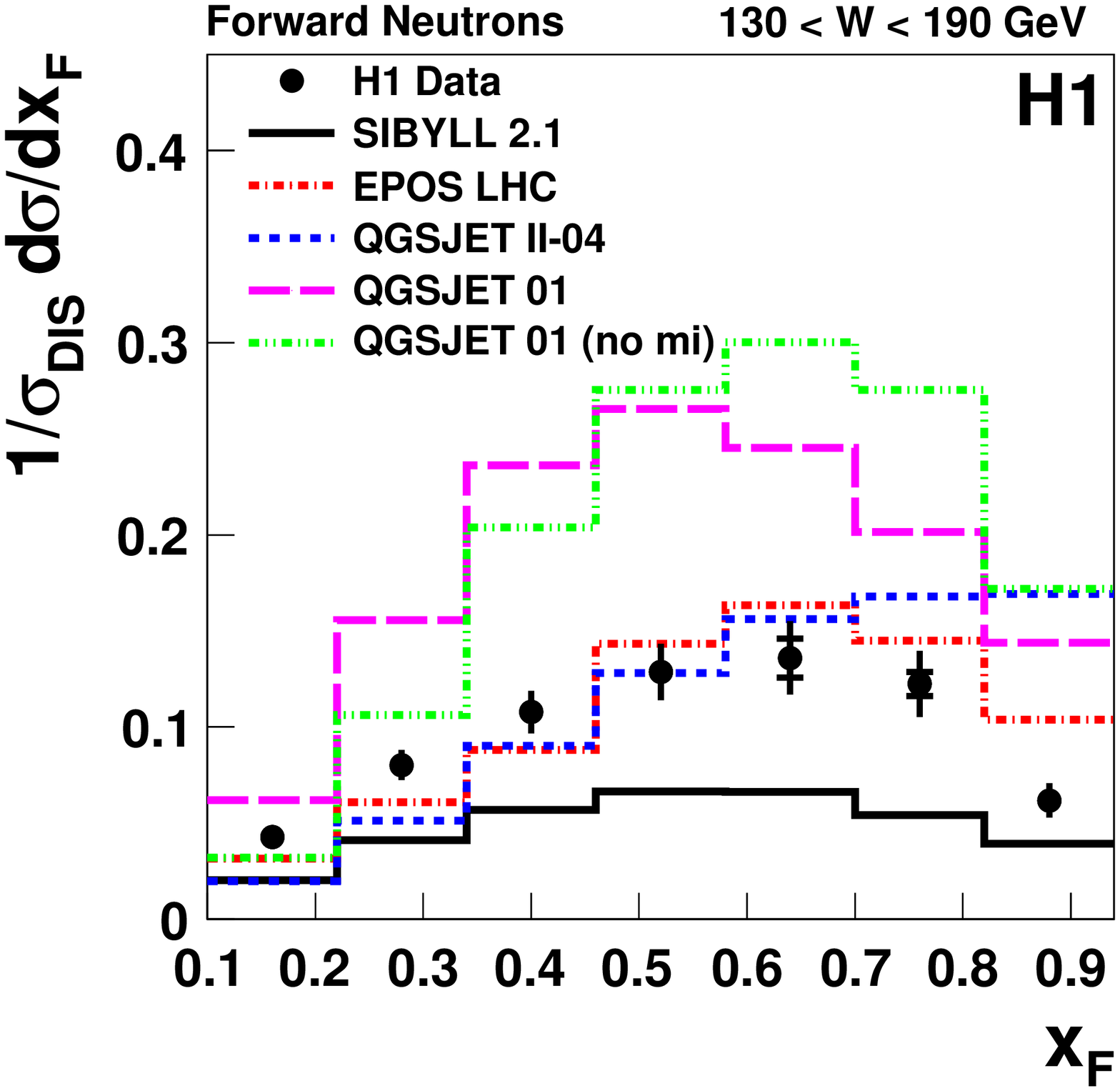,width=65mm}
\epsfig{file=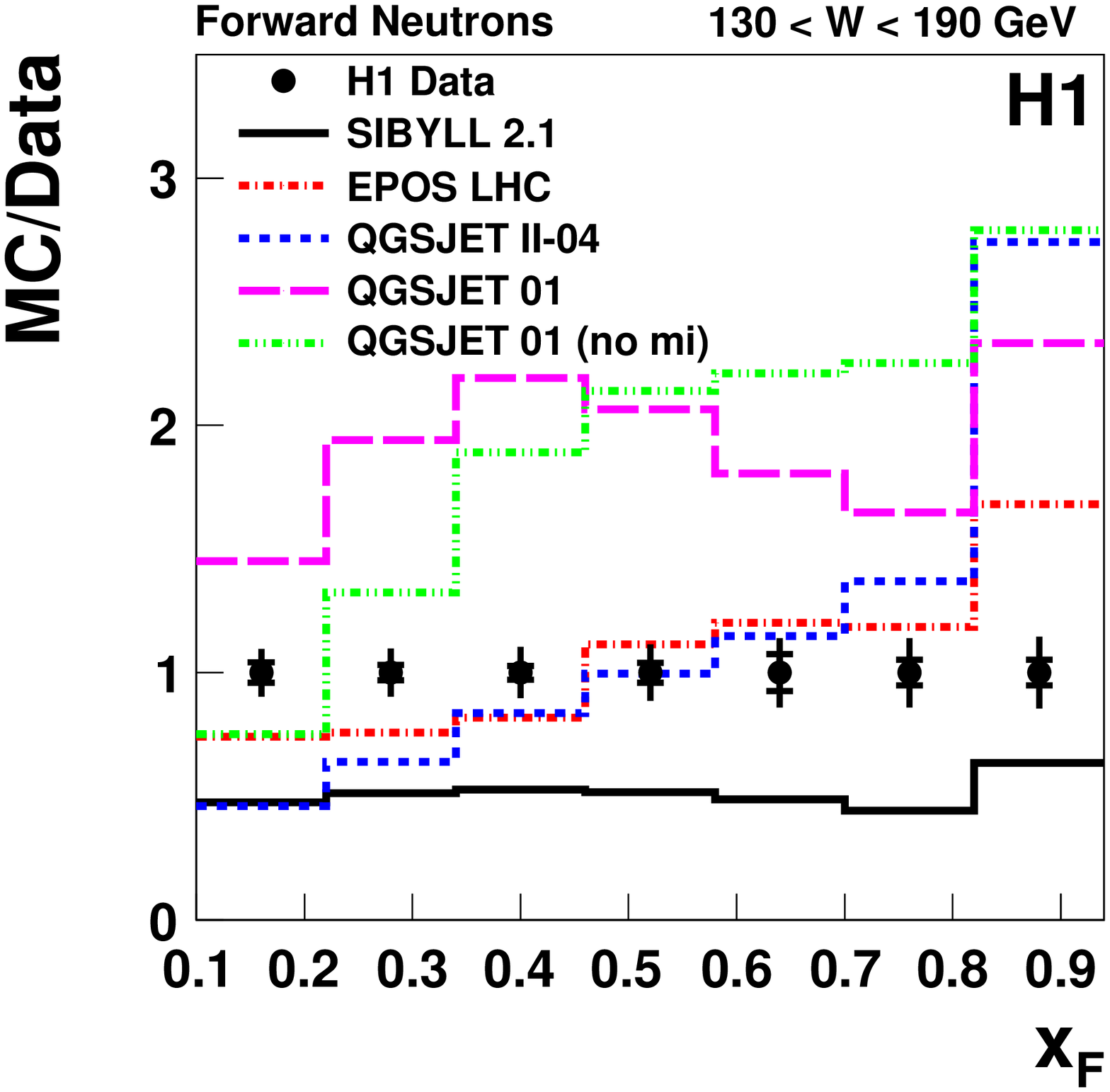,width=65mm}

\epsfig{file=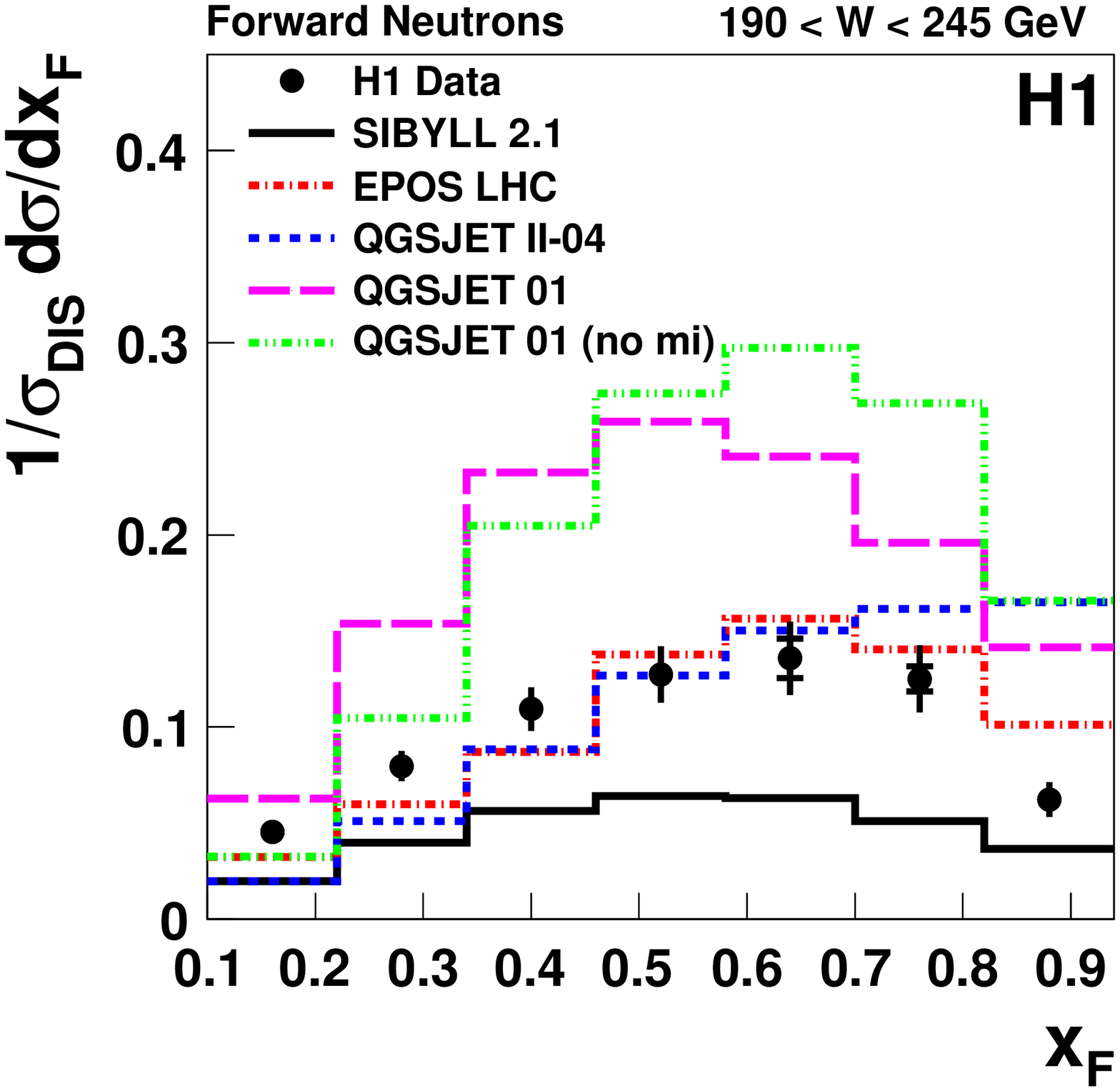,width=65mm}
\epsfig{file=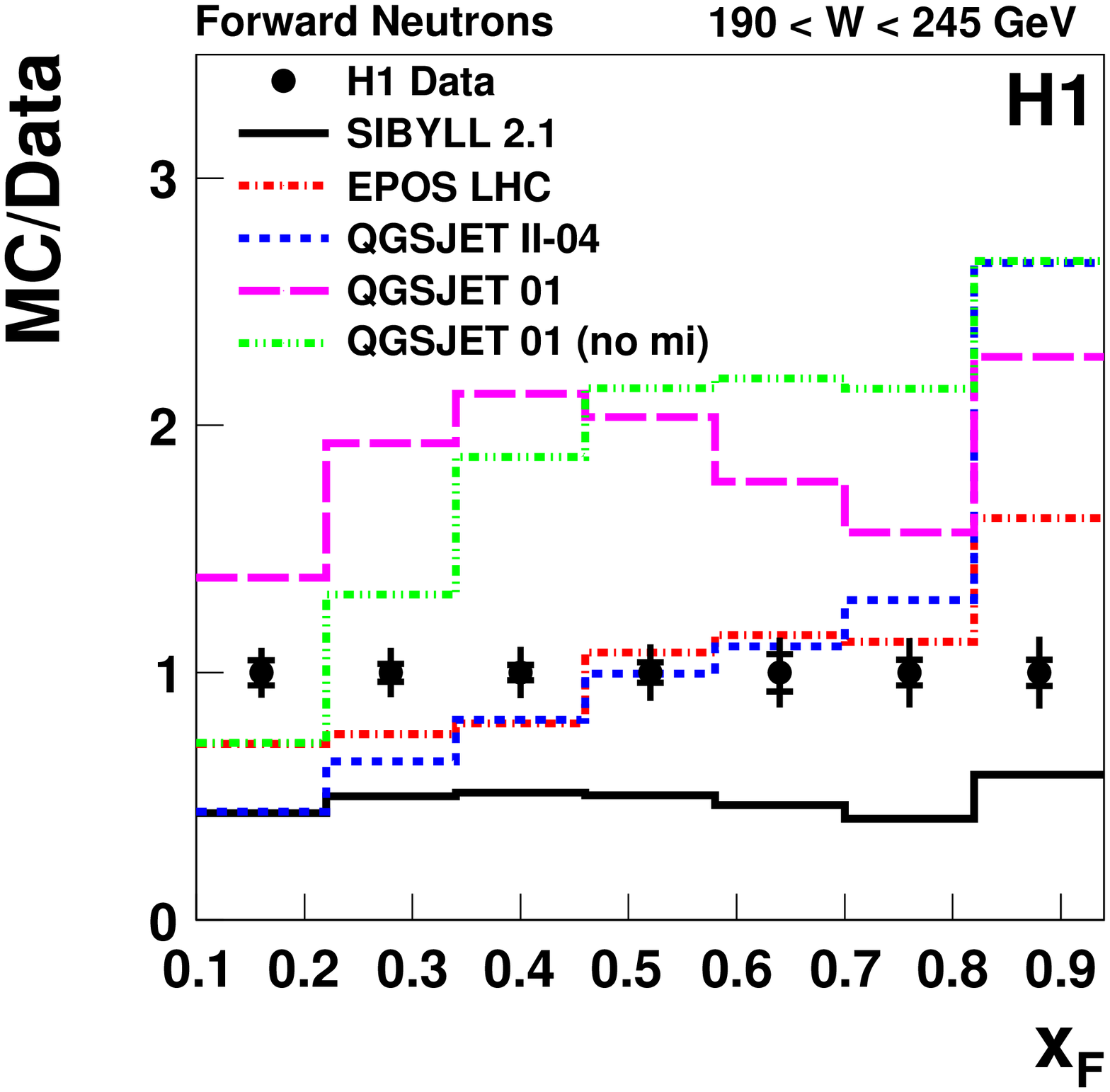,width=65mm}

\caption{
Normalised cross sections of forward neutron production in DIS
as a function of $x_F$ in three $W$ intervals
in the kinematic region given in Table~\ref{tab:selection}.
The inner error bars show the statistical uncertainty, while
the outer error bars show the total experimental uncertainty, 
calculated using the
quadratic sum of the statistical and systematic uncertainties.  
Also shown are the predictions of the cosmic ray hadronic interaction
models SIBYLL~2.1 (solid line), QGSJET~01 (dashed line), 
QGSJET~01~(no~mi) (dash-double dotted line),  
QGSJET~II-04 (dotted line) and EPOS~LHC (dash-dotted line).
In the right column the ratios of the CR model predictions to the data are shown.
}
\label{fig:xfncos}
\end{figure}

The normalised differential cross sections for forward neutrons
are presented in 
Table~\ref{tab:table4} and  in
Figures~\ref{fig:xfn} and \ref{fig:xfncos}
for the kinematic region defined in Table~\ref{tab:selection}.
The $x_F$ distributions 
are well reproduced 
by a combination of CDM and RAPGAP-$\pi$, using the weighting factors
and normalisation as described in section \ref{sec:resultW}.
The individual contributions of the two models are shown in 
Figure~\ref{fig:xfn} as well.
Fragmentation, as simulated by CDM,  dominates the neutron 
production at lower $x_F$, while the contribution from 
pion exchange becomes significant at $x_F\gsim 0.7$.
The combination of LEPTO and RAPGAP-$\pi$ (not shown) also provides a 
good description of the measurements for the three $W$ ranges.
In Figure~\ref{fig:xfncos} the predictions of the CR hadronic 
interaction models are compared to the
forward neutron production cross sections.
The EPOS~LHC model provides a reasonable description
of the neutron $x_F$ distributions, except at the highest $x_F$ values.
The SIBYLL~2.1 model describes the shape of the $x_F$ spectra but
fails in the absolute rate.
The QGSJET~II-04 model 
shows a harder
$x_F$ dependence, 
and QGSJET~01 predicts a much too high neutron rate.


A modified version of the 
QGSJET~01 model, denoted 'QGSJET 01~(no~mi)' \cite{ostapchenko2}, 
in which the contribution of multi-parton interactions is excluded
(see section \ref{sec:mc}), is also compared to the measurements.
When  multi-parton interactions are switched off, the predicted $x_F$ spectra 
become harder without improving the data description.

\begin{figure}[ht]
\epsfig{file=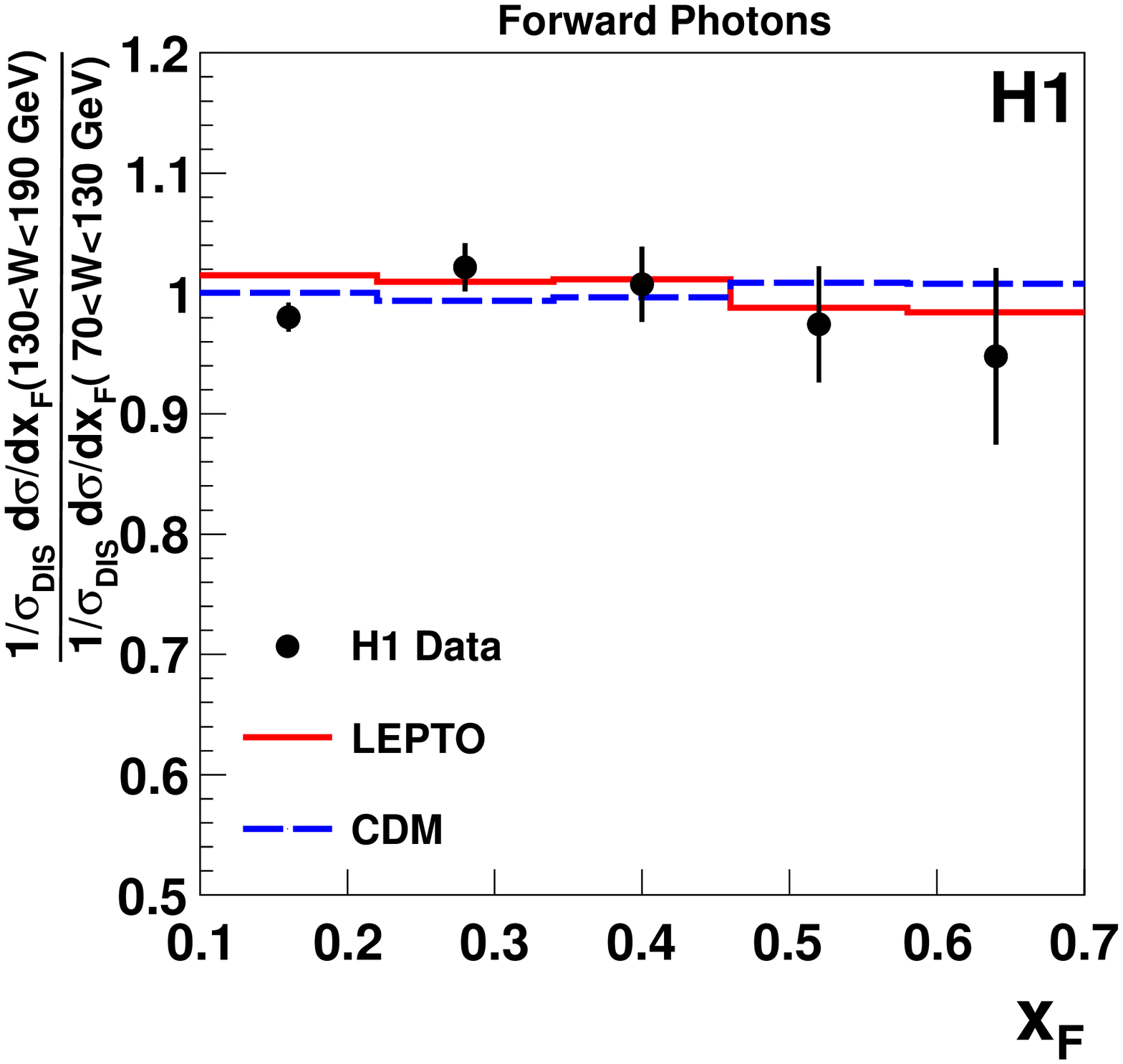,width=74mm}
\hspace*{5mm}
\epsfig{file=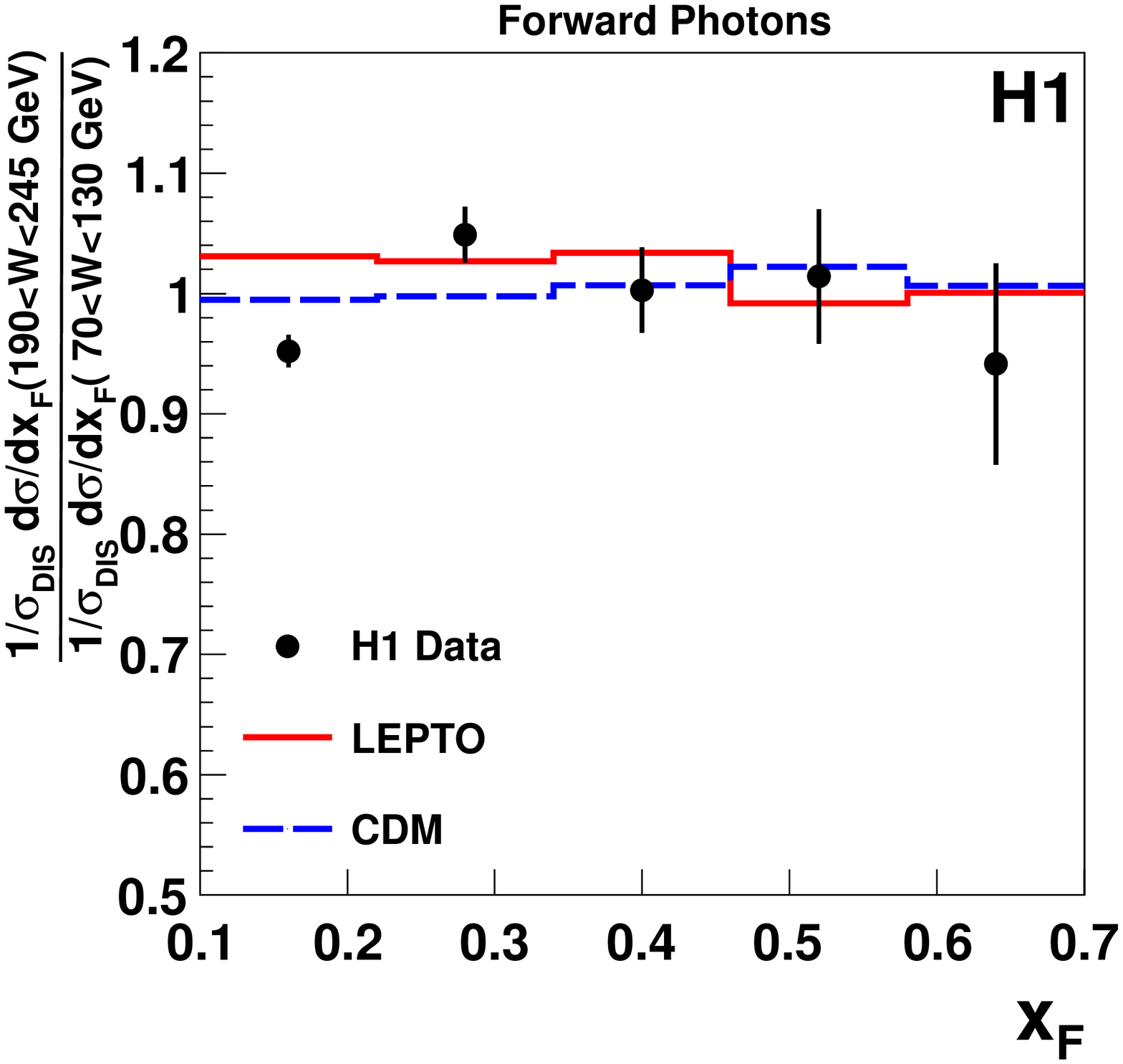,width=74mm}
\caption{
Ratios of normalised cross sections of forward photon production in DIS
corresponding to two different $W$ intervals, shown in Figure~\ref{fig:xfgmc},
 as a function of $x_F$:
{\bf (a)} ratio of the cross section in the $130<W<190$~GeV
interval to the cross section in the $70<W<130$~GeV interval;
{\bf (b)}   ratio of the cross section in the $190<W<245$~GeV
interval to the cross section in the $70<W<130$~GeV interval.
The kinematic phase space is defined in Table~\ref{tab:selection}.
The error bars show the total experimental uncertainty, calculated using the
quadratic sum of the statistical and systematic uncertainties.
Also shown are the predictions of the LEPTO (solid line) and 
CDM (dashed line) MC models.
}

\vspace*{-107mm}
{
\bf \large
\hspace*{17mm}(a)
\hspace*{74mm}(b)

}
\vspace*{102mm}
\label{fig:xfmcratio}
\end{figure}

\begin{figure}[h]
\epsfig{file=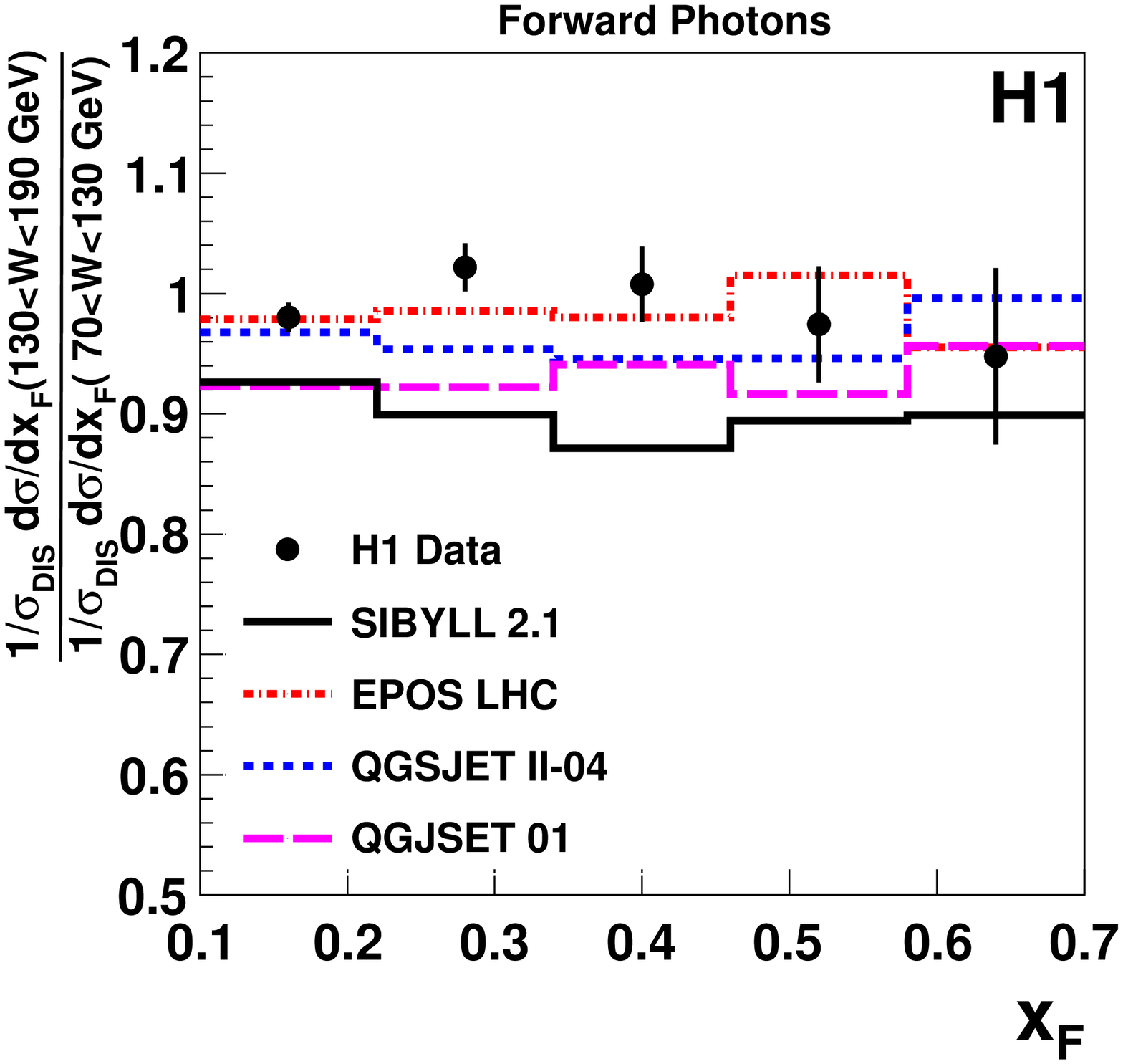,width=74mm}
\hspace*{5mm}
\epsfig{file=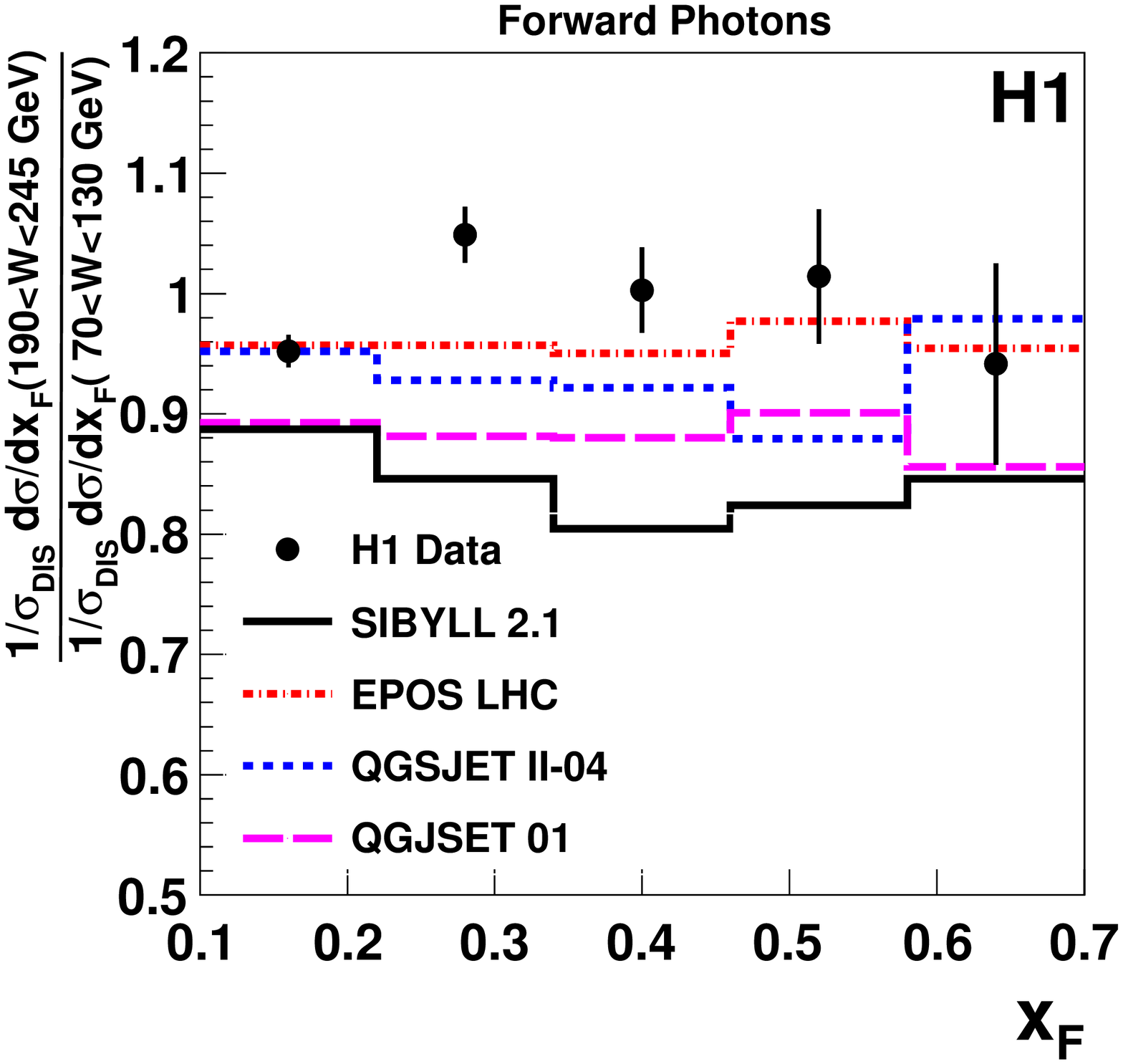,width=74mm}
\caption{
Ratios of normalised cross sections of forward photon production in DIS
corresponding to two different $W$ intervals, shown in Figure~\ref{fig:xfgcos}, 
as a function of $x_F$:
{\bf (a)} 
ratio of the cross section in the $130<W<190$~GeV
interval to the cross section in the $70<W<130$~GeV interval;
{\bf (b)} 
 ratio of the cross section in the $190<W<245$~GeV
interval to the cross section in the $70<W<130$~GeV interval.
The kinematic phase space is defined in Table~\ref{tab:selection}.
The error bars show the total experimental uncertainty, calculated using the
quadratic sum of the statistical and systematic uncertainties.
Also shown are the predictions of the cosmic ray hadronic interaction
models SIBYLL~2.1 (solid line), QGSJET~01 (dashed line),  
QGSJET~II-04 (dotted line) and EPOS~LHC (dash-dotted line).
}

\vspace*{-112mm}
{
\bf \large
\hspace*{17mm}(a)
\hspace*{74mm}(b)

}
\vspace*{110mm}
\label{fig:xfgcosratio}
\end{figure}

\begin{figure}[h]
\epsfig{file=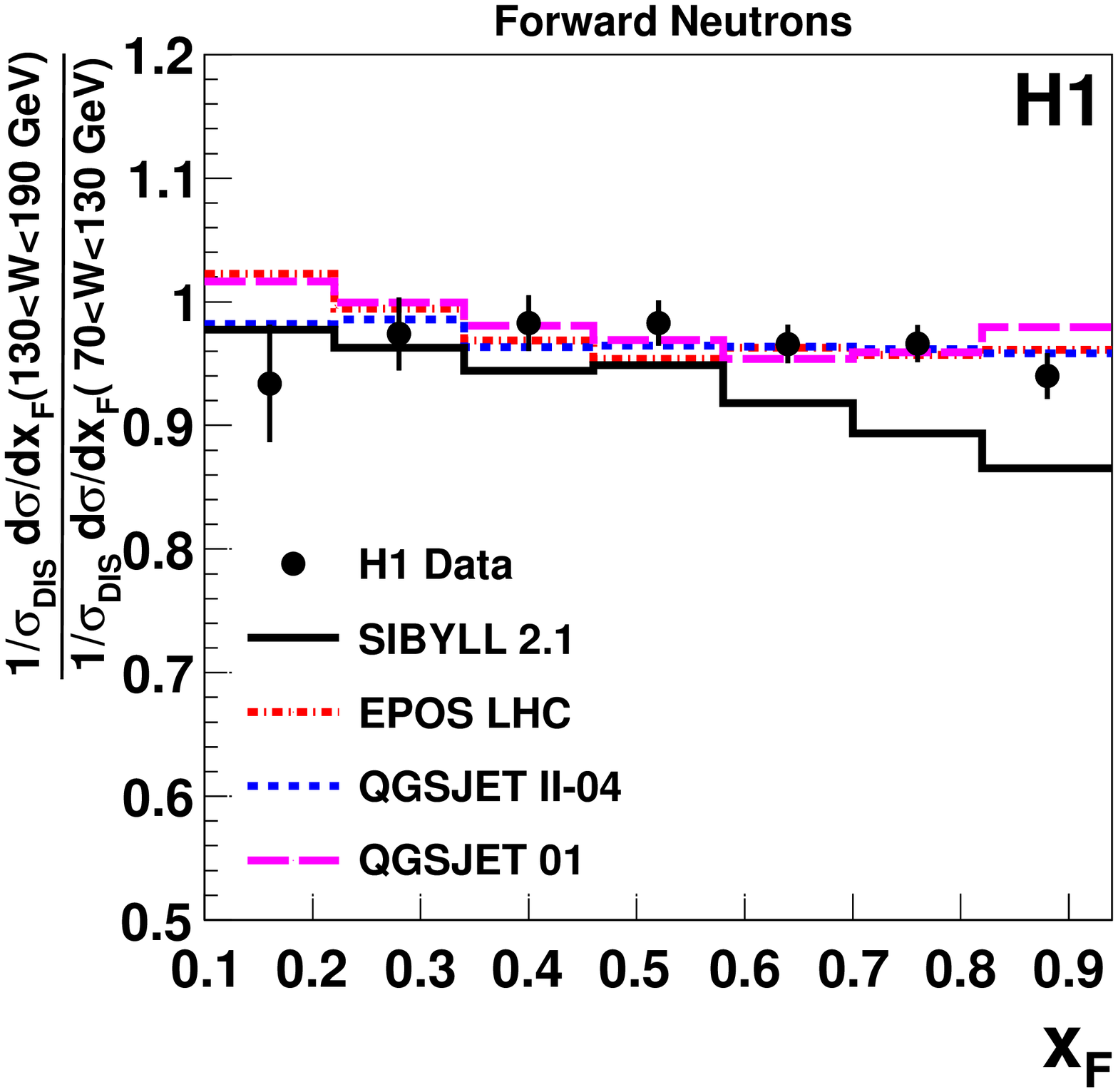,width=74mm}
\hspace*{5mm}
\epsfig{file=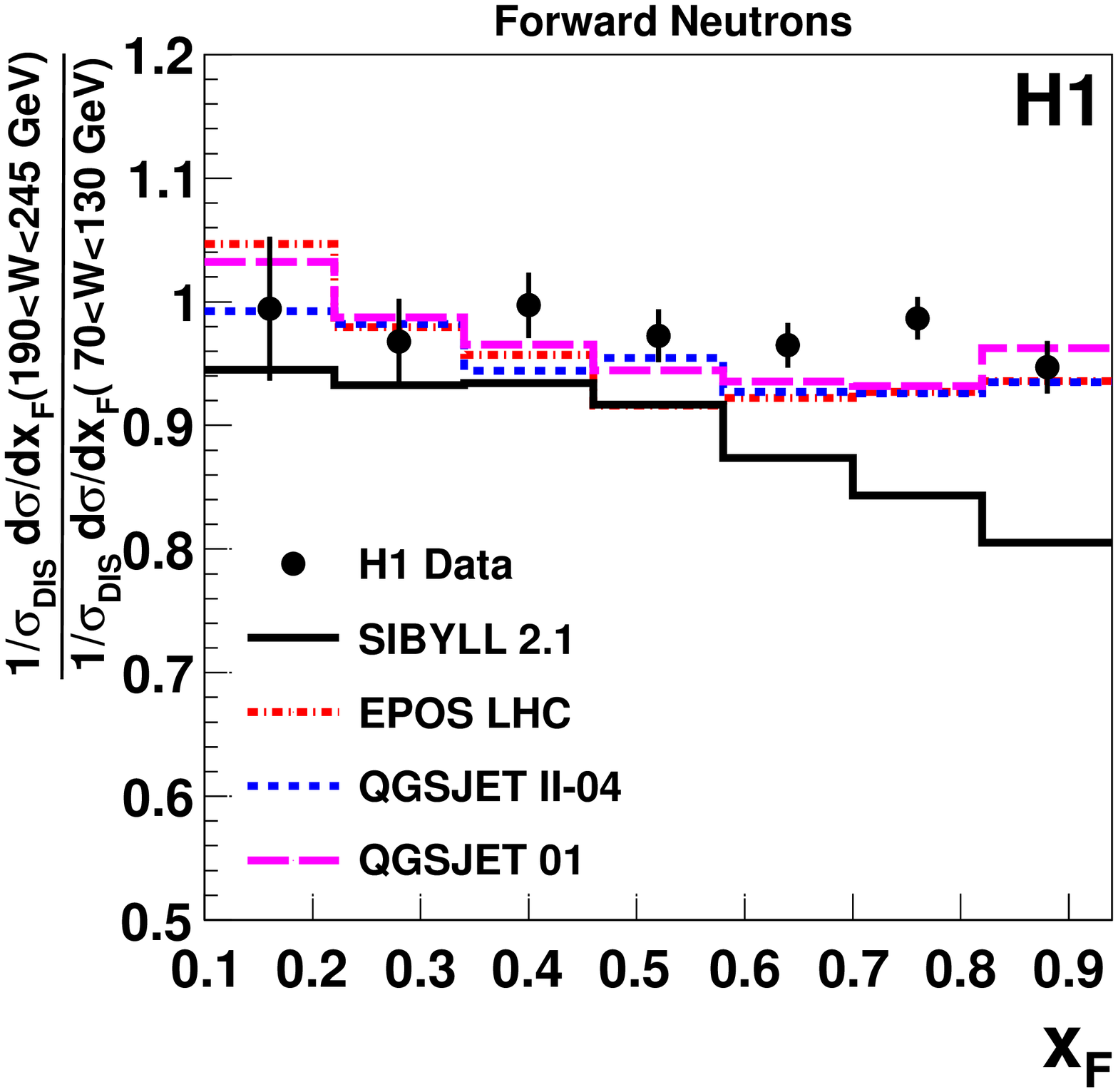,width=74mm}
\caption{
Ratios of normalised cross sections of forward neutron production in DIS
corresponding to two different $W$ intervals, shown in 
Figure~\ref{fig:xfncos},
as a function of $x_F$:
{\bf (a)}
ratio of the cross section in the $130<W<190$~GeV
interval to the cross section in the $70<W<130$~GeV interval;
{\bf (b)}
 ratio of the cross section in the $190<W<245$~GeV
interval to the cross section in the $70<W<130$~GeV interval.
The kinematic phase space is defined in Table~\ref{tab:selection}.
The error bars show the total experimental uncertainty, calculated using the
quadratic sum of the statistical and systematic uncertainties.
Also shown are the predictions of the cosmic ray hadronic interaction
models SIBYLL~2.1 (solid line), QGSJET~01 (dashed line),
QGSJET~II-04 (dotted line) and EPOS~LHC (dash-dotted line).
}
\vspace*{-117mm}
{
\bf \large
\hspace*{17mm}(a)
\hspace*{74mm}(b)

}
\vspace*{110mm}
\label{fig:xfnratio}
\end{figure}

The $W$ dependence of the $x_F$ distributions allows a test of the 
Feynman scaling hypothesis for particle production. For this test, the 
ratios of the normalised cross sections for different CM energy 
intervals are studied as a function of $x_F$. 
Figures~\ref{fig:xfmcratio}~and~\ref{fig:xfgcosratio} show the ratios of the second 
to the first and the third to the first $W$ range for photons.
The predictions from CDM, LEPTO and the CR models are also shown.
In Figure~\ref{fig:xfnratio} the same ratios are shown for forward neutrons and
the CR models are compared to the data.
For all data distributions the values of these ratios 
are 
consistent with unity and with being constant
within uncertainties, suggesting that Feynman scaling
in the target fragmentation region holds for photons and neutrons.
The LEPTO and CDM MC models, used for the comparison with forward
photon data, show a similar behaviour. 
All CR models indicate deviations from  scaling
for the forward photons, such that the production
rate decreases with increasing $W$. In particular,
this effect is strong for the SIBYLL~2.1 and QGSJET~01 models.
For forward neutrons the CR models
are consistent with Feynman scaling, with exception of SIBYLL~2.1.


\section{Summary}
\noindent
The production of high energy forward neutrons and 
photons has been studied at HERA
in deep-inelastic $ep$ scattering in the kinematic region
$6< Q^2 < 100~\GeV^2$ and $0.05<y<0.6$.
The normalised DIS cross sections $1/\sigma_{DIS}~{\rm d}\sigma/{\rm d}x_F$
for the production of  photons and neutrons at pseudorapidities 
$\eta>7.9$ and in the range of Feynman-$x$ of $0.1<x_F<0.7$ for the photons 
and $0.1<x_F<0.94$ for neutrons are presented.
The measured cross sections as a function of $x_F$ at 
different centre-of-mass
energies of the virtual photon-proton system agree within uncertainties,
confirming the validity of Feynman scaling in the energy range
of the virtual photon-proton system $70<W<245~\GeV$.

Different Monte Carlo models are compared to the measurements.
All these models overestimate the rate of photons by $30-70\%$.
The shapes of the measured forward photon 
cross sections are well described by the LEPTO
MC simulation, while predictions based on the colour dipole model fail,
especially at high $x_F$. 
The cross sections for forward neutrons are well described by 
a linear combination of the standard fragmentation model, as implemented in
DJANGOH, and 
the one-pion-exchange model RAPGAP-$\pi$.
Predictions of models,  which are commonly used for the simulation of
cosmic ray cascades, are also compared to the 
forward photon and neutron measurements.
None of the models describes the photon and neutron data 
simultaneously well. The best description of the shapes
of the photon and the neutron $x_F$ distributions is
provided by the EPOS~LHC model.
Within the kinematic range of the measurements, 
the relative rate of forward photons
and neutrons in DIS events is observed to be independent of the energy
of the virtual photon-proton CM, 
and therefore also consistent
with the hypothesis of limiting fragmentation.
The present measurement provides new information to further improve
the understanding of proton fragmentation in collider and 
cosmic ray experiments.

\section*{Acknowledgements}

We are grateful to the HERA machine group whose outstanding
efforts have made this experiment possible. 
We thank the engineers and technicians for their work in constructing and
maintaining the H1 detector, our funding agencies for 
financial support, the
DESY technical staff for continual assistance
and the DESY directorate for support and for the
hospitality which they extend to the non DESY 
members of the collaboration.
We would like to give credit to all partners contributing to the EGI 
computing infrastructure for their support for the H1 Collaboration.
We also wish to thank Tanguy Pierog, Ralph Engel and Sergey Ostapchenko
for providing the predictions of the cosmic ray models and for 
fruitful discussions.

\newpage

\begin{thebibliography}{10}

\bibitem{Engel:1998hf}
R.~Engel,
\newblock Nucl. Phys. Proc. Suppl. {\bf 75A} (1999) 62 [astro-ph/9811225].

\bibitem{Bunyatyan:2009zz}
A.~Buniatyan {\em et~al.},
{\it ''Cosmic Rays, HERA and the LHC: Working group summary."}
Proceedings of the Workshop on the Implications of HERA for LHC Physics,
Geneva, Switzerland, 26-30 May 2008, p.566, DESY-PROC-2009-02 
[arXiv:0903.3861].

\bibitem{Adloff:1998yg}
C.~Adloff {\em et~al.} [H1 Collaboration],
\newblock Eur. Phys. J. {\bf C6} (1999) 587 [hep-ex/9811013].

\bibitem{Chekanov:2002pf}
S.~Chekanov {\em et~al.} [ZEUS Collaboration],
\newblock Nucl. Phys. {\bf B637} 3 (2002) 3  [hep-ex/0205076].

\bibitem{Chekanov:2007tv}
S.~Chekanov {\em et~al.} [ZEUS Collaboration],
\newblock Nucl. Phys. {\bf B776} (2007)  1 [hep-ex/0702028].

\bibitem{Chekanov:2008tn}
S.~Chekanov {\em et~al.} [ZEUS Collaboration],
\newblock JHEP {\bf 06} (2009) 074 [arXiv:0812.2416].

\bibitem{Aaron:2010ab}
F.~Aaron {\em et~al.} [H1 Collaboration],
\newblock Eur. Phys. J. {\bf C68} (2010) 381 [arXiv:1001.0532].

\bibitem{Aaron:2011pe}
F.~Aaron {\em et~al.} [H1 Collaboration],
\newblock Eur. Phys. J. {\bf C71} (2011) 1771 [arXiv:1106.5944].

\bibitem{Benecke:1969sh}
J.~Benecke  {\em et~al.}, 
\newblock Phys. Rev. {\bf 188} (1969) 2159.

\bibitem{Chou:1994dh}
T.~T. Chou and C.-N. Yang,
\newblock Phys. Rev. {\bf D50} (1994) 590.

\bibitem{Feynman:1969ej}
R.~P. Feynman,
\newblock Phys. Rev. Lett. {\bf 23} (1969) 1415.

\bibitem{Arnison:1982ed}
G.~Arnison {\em et~al.} [UA1 Collaboration],
\newblock Phys. Lett. {\bf B118} (1982) 167.

\bibitem{Arnison:1982rm}
G.~Arnison {\em et~al.} [UA1 Collaboration],
\newblock Phys. Lett. {\bf B123} (1983) 108.

\bibitem{Banner:1984wh}
M.~Banner {\em et~al.} [UA2 Collaboration],
\newblock Z. Phys. {\bf C27} (1985) 329.

\bibitem{Bernard:1985kh}
D.~Bernard {\em et~al.} [UA4 Collaboration],
\newblock Phys. Lett. {\bf B166} (1986) 459.

\bibitem{Alner:1986xu}
G.~Alner {\em et~al.} [UA5 Collaboration],
\newblock Z. Phys. {\bf C33} (1986) 1.

\bibitem{Alner:1987wb}
G.~Alner {\em et~al.} [UA5 Collaboration],
\newblock Phys. Rep. {\bf 154} (1987) 247.

\bibitem{Alexopoulos:1988na}
T.~Alexopoulos {\em et~al.},
\newblock Phys. Rev. Lett. {\bf 60} (1988) 1622.

\bibitem{Abe:1989td}
F.~Abe {\em et~al.} [CDF Collaboration],
\newblock Phys. Rev. {\bf D41} (1990) 2330.

\bibitem{Aaron:2007ds}
F.~Aaron {\em et~al.} [H1 Collaboration],
\newblock Phys. Lett. {\bf B654} (2007) 148 [arXiv:0706.2456].

\bibitem{Abramowicz:2010rz}
H.~Abramowicz {\em et~al.} [ZEUS Collaboration],
\newblock JHEP {\bf 1006} (2010) 009 [arXiv:1001.4026].

\bibitem{Pare:1989mr}
E.~Pare {\em et~al.} [UA7 Collaboration],
\newblock Phys. Lett. {\bf B242} (1990) 531.

\bibitem{Albrow:1973kj}
M.~Albrow {\em et~al.} [CHLM Collaboration],
\newblock Nucl. Phys. {\bf B56} (1973) 333.

\bibitem{Albrow:1973ee}
M.~Albrow {\em et~al.} [CHLM Collaboration],
\newblock Nucl. Phys. {\bf B73} (1974) 40.

\bibitem{Capiluppi:1973fz}
P.~Capiluppi {\em et~al.},
\newblock Nucl. Phys. {\bf B70} (1974) 1.

\bibitem{Capiluppi:1974rt}
P.~Capiluppi {\em et~al.},
\newblock Nucl. Phys. {\bf B79} (1974) 189.

\bibitem{Ohsawa:1992dj}
A.~Ohsawa, 
\newblock Prog. Theor. Phys. {\bf 92} (1994) 1005.

\bibitem{Abt:1996hi}
I.~Abt {\em et~al.} [H1 Collaboration],
\newblock Nucl. Instrum. Meth. {\bf A386} (1997) 310.

\bibitem{Abt:1996xv}
I.~Abt {\em et~al.} [H1 Collaboration],
\newblock Nucl. Instrum. Meth. {\bf A386} (1997) 348.

\bibitem{Appuhn:1996na}
R.~D. Appuhn {\em et~al.} [H1 SPACAL Group],
\newblock Nucl. Instrum. Meth. {\bf A386} (1997) 397.

\bibitem{Pitzl:2000wz}
D.~Pitzl {\em et~al.},
\newblock Nucl. Instrum. Meth. {\bf A454} (2000) 334 [hep-ex/0002044].

\bibitem{Andrieu:1993kh}
B.~Andrieu {\em et~al.} [H1 Calorimeter Group],
\newblock Nucl. Instrum. Meth. {\bf A336} (1993) 460.

\bibitem{Laycock:2012xg}
P.J.~Laycock {\em et~al.},
\newblock JINST {\bf 7} (2012) T08003 [arXiv:1206:4068].


\bibitem{Nicholls:1995di}
T.~Nicholls {\em et al.} [H1 SPACAL Group],
\newblock Nucl. Instrum. Meth. {\bf A374} (1996)  149.

\bibitem{Aaron:2012kn}
F.~Aaron {\em et al.} [H1 Collaboration], 
Eur. Phys. J. {\bf C72} (2012) 2163, Erratum-ibid. {\bf C74} (2014) 2733
 [arXiv:1205.2448].

\bibitem{Uraev}
A.~Uraev, 
  ``{\it A new Forward Neutron Calorimeter for the H1 Experiment at the HERA collider.}" 
  (In Russian), Diploma thesis, Moscow Engineering Physics Institute (2001) 
  (available at  http://www-h1.desy.de/publications/theses\_list.html).

\bibitem{Peez:2003zd}
M.~Peez,
``{\it Search for deviations from the standard model in high transverse energy
  processes at the electron proton collider HERA.}" (In French), PhD thesis,
  Univ.~Lyon (2003), CPPM-T-2003-04 
  (available at http://www-h1.desy.de/publications/theses\_list.html).

\bibitem{Hellwig:2004yp}
 S.~Hellwig, ``{\it Investigation of the $D^* - \pi_{slow}$ double tagging method in
  charm analyses.}" (In German), Diploma thesis, Univ.~Hamburg (2004) 
 (available at http://www-h1.desy.de/publications/theses\_list.html).

\bibitem{Adloff:1997sc}
C.~Adloff {\em et~al.} [H1 Collaboration],
\newblock Z. Phys. {\bf C76} (1997) 613 [hep-ex/9708016].

\bibitem{Brun:1978fy}
R.~Brun  {\em et~al.}, GEANT3, CERN-DD/EE/84-1.

\bibitem{Charchula:1994kf}
K.~Charchula, G.~A. Schuler and H.~Spiesberger,
\newblock DJANGOH~1.4,
\newblock Comput. Phys. Commun. {\bf 81} (1994) 381.

\bibitem{Ingelman:1996mq}
G.~Ingelman, A.~Edin and J.~Rathsman,
\newblock LEPTO 6.5,
\newblock Comput. Phys. Commun. {\bf 101} (1997) 108 [hep-ph/9605286].

\bibitem{Lonnblad:1992tz}
L.~L\"onnblad,
\newblock ARIADNE~4.10,
\newblock Comput. Phys. Commun. {\bf 71} (1992) 15.

\bibitem{Andersson:1983ia}
B.~Andersson  {\em et~al.},
\newblock Phys. Rep. {\bf 97} (1983) 31.

\bibitem{Sjostrand:1995iq}
T.~Sj\"ostrand,
\newblock PYTHIA 5.7 and JETSET 7.4,
\newblock hep-ph/9508391.

\bibitem{Kwiatkowski:1990es}
A.~Kwiatkowski, H.~Spiesberger and H.~J. M\"ohring,
\newblock Comp. Phys. Commun. {\bf 69} (1992) 155.

\bibitem{Edin:1995gi}
A.~Edin, G.~Ingelman and J.~Rathsman,
\newblock Phys. Lett. {\bf B366} (1996) 371 [hep-ph/9508386].

\bibitem{Aaron:2009kv}
F.~D. Aaron {\em et~al.} [H1 Collaboration],
\newblock Eur. Phys. J. {\bf C64} (2009) 561 [arXiv:0904.3513].

\bibitem{Jung:1993gf}
H.~Jung,
\newblock RAPGAP 3.1,
\newblock Comp. Phys. Commun. {\bf 86} (1995) 147.

\bibitem{Bishari:1972tx}
M.~Bishari,
\newblock Phys. Lett. {\bf B38} (1972) 510.

\bibitem{Holtmann:1994rs}
H.~Holtmann {\em et~al.},
\newblock Phys. Lett. {\bf B338} (1994) 363.

\bibitem{Kopeliovich:1996iw}
B.~Kopeliovich, B.~Povh and I.~Potashnikova,
\newblock Z. Phys. {\bf C73} (1996) 125 [hep-ph/9601291].

\bibitem{Przybycien:1996zb}
M.~Przybycie\'n, A.~Szczurek and G.~Ingelman,
\newblock Z. Phys. {\bf C74} (1997) 509 [hep-ph/9606294].

\bibitem{Szczurek:1997cw}
A.~Szczurek, N.~N. Nikolaev and J.~Speth,
\newblock Phys. Lett. {\bf B428} (1998) 383 [hep-ph/9712261].

\bibitem{Werner:2005jf}
K.~Werner, F.-M. Liu and T.~Pierog,
\newblock Phys. Rev. {\bf C74} (2006) 044902 [hep-ph/0506232].

\bibitem{Pierog:2013ria}
T.~Pierog  {\em et~al.},
\newblock DESY-13-125, arXiv:1306.0121.

\bibitem{Kalmykov:1993qe}
N.~N. Kalmykov and S.~S. Ostapchenko,
\newblock Phys. Atom. Nucl. {\bf 56} (1993) 346.

\bibitem{Kalmykov:1997te}
N.~N. Kalmykov, S.~S. Ostapchenko and A.~I. Pavlov,
\newblock Nucl. Phys. Proc. Suppl. {\bf 52B} (1997) 17.

\bibitem{Ostapchenko:2005nj}
S.~S. Ostapchenko,
\newblock Phys. Rev. {\bf D74} (2006) 014026 [hep-ph/0505259].

\bibitem{Ostapchenko:2010vb}
S.~S.~Ostapchenko,
\newblock Phys. Rev. {\bf D83} (2011) 014018 [arXiv:1010.1869].

\bibitem{Engel:1992vf}
J.~Engel  {\em et~al.},
\newblock Phys. Rev. {\bf D46} (1992) 5013.

\bibitem{Ahn:2009wx}
E.-J. Ahn  {\em et~al.},
\newblock Phys. Rev. {\bf D80} (2009) 094003 [arXiv:0906.4113].


\bibitem{Collins:1977jy}
P.~D.~B. Collins, ``{\it An Introduction to Regge Theory and High-Energy  Physics}.''
 Cambridge 1977, 445p.

\bibitem{Gribov:1968fc}
V.~N. Gribov,
\newblock Sov. Phys. JETP {\bf 26} (1968) 414.

\bibitem{Engel:1995yda}
R.~Engel and J.~Ranft,
\newblock Phys. Rev. {\bf D54} (1996) 4244 [hep-ph/9509373].

\bibitem{ostapchenko1}
S.~S. Ostapchenko,
\newblock {\em {Private communication}}.

\bibitem{ostapchenko2}
S.~S. Ostapchenko and T.~Pierog,
\newblock {\em {Private communication}}.

\bibitem{Adriani:2011nf}
O.~Adriani {\em et~al.} [LHCf Collaboration],
\newblock Phys. Lett. {\bf B703} (2011) 128  [arXiv:1104.5294].

\bibitem{Adriani:2012ap}
O.~Adriani {\em et~al.} [LHCf Collaboration],
\newblock Phys. Rev. {\bf D86} (2012) 092001  [arXiv:1205.4578].

\bibitem{Adriani:2012fz}
O.~Adriani {\em et~al.} [LHCf Collaboration],
\newblock Phys. Lett. {\bf B715} (2012) 298  [arXiv:1207.7183].



\end{thebibliography}

\newpage

\begin{table}[hp]
\centering
{
\small
\begin{tabular}{|c|c|c|c||c|c|c|c|}
\hline 
& & & & & \multicolumn{3}{|c|}{\underline{\hspace*{0.1cm} correlated sys. uncertainty\hspace*{0.1cm}} } \\
$W$ range $[GeV]$  &  { \Large$\frac{\sigma^\gamma_{DIS}(W)}{\sigma_{DIS}(W)}$} & $\delta_{stat.}$ & 
$\delta_{total~sys.}$ &  $\delta_{uncorrel.sys.}$  &  
$\delta_{E_{FNC}}$ & $\delta_{XY_{FNC}}$ & $\delta_{model}$   \\
&  & & & & & & \\ [-5pt]
\hline 
&  & & & & & & \\ [-12pt]
$  ~70.  \div 115.$ & $0.0269$  &  $0.0002$ & $0.0022$ &  $0.0006$ & $0.0011$ & $0.0018$ & $0.0003$ \\ 
$  115.  \div 160.$ & $0.0269$  &  $0.0002$ & $0.0022$ &  $0.0007$ & $0.0011$ & $0.0018$ & $0.0003$ \\
$  160.  \div 205.$ & $0.0265$  &  $0.0002$ & $0.0022$ &  $0.0007$ & $0.0011$ & $0.0018$ & $0.0003$ \\
$  205.  \div 245.$ & $0.0265$  &  $0.0002$ & $0.0022$ &  $0.0007$ & $0.0011$ & $0.0018$ & $0.0003$ \\
\hline
\end{tabular}
\normalsize

}
\caption{
The fraction of DIS events with forward photons
in the kinematic region given in Table~\ref{tab:selection}.
For each measurement, the  statistical ($\delta_{stat.}$), 
the total systematic  ($\delta_{total~sys.}$), the
uncorrelated  ($\delta_{uncorrel.sys.}$)
systematic uncertainties and the bin-to-bin correlated
systematic uncertainties due to the FNC absolute 
energy scale  ($\delta_{E_{FNC}}$), 
the measurement of the particle impact position 
in the FNC  ($\delta_{XY_{FNC}}$)
and the model dependence of the data  
correction ($\delta_{model}$) are given.
}
\label{tab:table1}
\end{table}

\begin{table}[hp]
\centering
{
\small
\begin{tabular}{|c|c|c|c||c|c|c|c|}
\hline
& & & & & \multicolumn{3}{|c|}{\underline{\hspace*{0.1cm} correlated sys. uncertainty\hspace*{0.1cm}} } \\
$W$ range $[GeV]$  &  { \Large$\frac{\sigma^n_{DIS}(W)}{\sigma_{DIS}(W)}$} & $\delta_{stat.}$ & 
$\delta_{total~sys.}$ &  $\delta_{uncorrel.sys.}$  &  
$\delta_{E_{FNC}}$ & $\delta_{XY_{FNC}}$ & $\delta_{model}$   \\
&  & & & & & & \\ [-5pt]
\hline
&  & & & & & & \\ [-12pt]
$  ~70.  \div 115.$ & $0.0843 $  &  $0.0004$ & $0.0074$ &  $0.0020$ & $0.0008$ & $0.0057$ & $0.0042$ \\ 
$  115.  \div 160.$ & $0.0830 $  &  $0.0004$ & $0.0074$ &  $0.0021$ & $0.0008$ & $0.0056$ & $0.0042$ \\
$  160.  \div 205.$ & $0.0815 $  &  $0.0005$ & $0.0072$ &  $0.0020$ & $0.0008$ & $0.0055$ & $0.0041$ \\
$  205.  \div 245.$ & $0.0826 $  &  $0.0006$ & $0.0073$ &  $0.0022$ & $0.0008$ & $0.0055$ & $0.0041$ \\
\hline
\end{tabular}
\normalsize

}
\caption{
The fraction of DIS events with forward neutrons
in the kinematic region given in Table~\ref{tab:selection}.
For each measurement, the  statistical ($\delta_{stat.}$), 
the total systematic  ($\delta_{total~sys.}$), the
uncorrelated  ($\delta_{uncorrel.sys.}$)
systematic uncertainties and the bin-to-bin correlated
systematic uncertainties due to the FNC absolute 
energy scale  ($\delta_{E_{FNC}}$), 
the measurement of the particle impact position 
in the FNC  ($\delta_{XY_{FNC}}$)
and the model dependence of the data
correction ($\delta_{model}$) are given.
}
\label{tab:table2}
\end{table}

\begin{table}[hp]
\centering
{
\small
\begin{tabular}{|c|l|l|l||l|l|l|l|}
\multicolumn{8}{c}{$W=70-130~\rm GeV$} \\
\hline  
& & & & & \multicolumn{3}{|c|}{\underline{\hspace*{0.1cm} correlated sys. uncertainty\hspace*{0.1cm}} } \\
$x_F$ range  &  {\Large$\frac{1}{\sigma_{DIS}}\frac{d\sigma}{d x_F}$} & 
$\delta_{stat.}$ & $\delta_{total~sys.}$ &  $\delta_{uncorrel.sys.}$  &  
$\delta_{E_{FNC}}$ & $\delta_{XY_{FNC}}$ & $\delta_{model}$   \\
& & & & &  & & \\ [-5pt]
\hline 
&  & & & & & & \\ [-12pt]
$ 0.10 \div 0.22 $ &  \hspace{5mm}  $0.130$   & $0.001$  & $0.011$   & $0.003$  & $0.001$  & $0.011$   & $0.001$   \\ 
$ 0.22 \div 0.34 $ &  \hspace{5mm}  $0.0542$  & $0.0007$ & $0.0060$  & $0.0015$ & $0.0027$ & $0.0051$  & $0.0008$  \\
$ 0.34 \div 0.46 $ &  \hspace{5mm}  $0.0221$  & $0.0005$ & $0.0031$  & $0.0007$ & $0.0018$ & $0.0024$  & $0.0003$  \\
$ 0.46 \div 0.58 $ &  \hspace{5mm}  $0.00743$ & $0.00024$ & $0.00122$  & $0.00032$ & $0.00059$ & $0.00099$  & $0.00026$  \\
$ 0.58 \div 0.70 $ &  \hspace{5mm}  $0.00202$ & $0.00010$ & $0.00044$  & $0.00016$ & $0.00022$ & $0.00031$  & $0.00014$  \\ 
\hline
\end{tabular}
\vspace*{4mm}

\begin{tabular}{|c|l|l|l||l|l|l|l|}
\multicolumn{8}{c}{$W=130-190~\rm GeV$} \\
\hline  
& & & & & \multicolumn{3}{|c|}{\underline{\hspace*{0.1cm} correlated sys. uncertainty\hspace*{0.1cm}} } \\
$x_F$ range  &  {\Large$\frac{1}{\sigma_{DIS}}\frac{d\sigma}{d x_F}$} & 
$\delta_{stat.}$ & $\delta_{total~sys.}$ &  $\delta_{uncorrel.sys.}$  & 
$\delta_{E_{FNC}}$ & $\delta_{XY_{FNC}}$ & $\delta_{model}$   \\
& & & & & & &  \\ [-12pt]
& & & & & & &  \\ [-8pt]
\hline
& & &  & & & & \\ [-12pt]
$ 0.10 \div 0.22 $ &  \hspace{5mm}  $0.128$   & $0.001$  & $0.011$   & $0.003$  & $0.001$  & $0.011$   & $0.001$   \\ 
$ 0.22 \div 0.34 $ &  \hspace{5mm}  $0.0553$  & $0.0008$ & $0.0063$  & $0.0016$ & $0.0028$ & $0.0053$  & $0.0008$  \\
$ 0.34 \div 0.46 $ &  \hspace{5mm}  $0.0222$  & $0.0005$ & $0.0031$  & $0.0007$ & $0.0018$ & $0.0024$  & $0.0003$  \\
$ 0.46 \div 0.58 $ &  \hspace{5mm}  $0.00724$ & $0.00027$ & $0.00120$  & $0.00032$ & $0.00058$ & $0.00097$  & $0.00025$  \\
$ 0.58 \div 0.70 $ &  \hspace{5mm}  $0.00192$ & $0.00011$ & $0.00041$  & $0.00015$ & $0.00021$ & $0.00029$  & $0.00013$  \\ 
\hline
\end{tabular}
\vspace*{4mm}

\begin{tabular}{|c|l|l|l||l|l|l|l|}
\multicolumn{8}{c}{$W=190-245~\rm GeV$} \\
\hline
& & & & & \multicolumn{3}{|c|}{\underline{\hspace*{0.1cm} correlated sys. uncertainty\hspace*{0.1cm}} } \\
$x_F$ range  &  {\Large$\frac{1}{\sigma_{DIS}}\frac{d\sigma}{d x_F}$} & 
$\delta_{stat.}$ & $\delta_{total~sys.}$ &  $\delta_{uncorrel.sys.}$  & 
$\delta_{E_{FNC}}$ & $\delta_{XY_{FNC}}$ & $\delta_{model}$  \\
& & & & & & &   \\ [-8pt]
\hline 
&  & & & & & & \\ [-12pt]
$ 0.10 \div 0.22 $ &  \hspace{5mm}  $0.124$   & $0.001$  & $0.011$   & $0.003$  & $0.001$  & $0.011$   & $0.001$   \\ 
$ 0.22 \div 0.34 $ &  \hspace{5mm}  $0.0568$  & $0.0010$ & $0.0064$  & $0.0017$ & $0.0028$ & $0.0054$  & $0.0008$  \\
$ 0.34 \div 0.46 $ &  \hspace{5mm}  $0.0222$  & $0.0006$ & $0.0031$  & $0.0007$ & $0.0018$ & $0.0024$  & $0.0003$  \\
$ 0.46 \div 0.58 $ &  \hspace{5mm}  $0.00754$ & $0.00034$ & $0.00125$  & $0.00033$ & $0.00060$ & $0.00101$  & $0.00026$  \\
$ 0.58 \div 0.70 $ &  \hspace{5mm}  $0.00190$ & $0.00014$ & $0.00041$  & $0.00015$ & $0.00021$ & $0.00029$  & $0.00013$  \\ 
\hline
\end{tabular}
\normalsize

}
\caption{
Normalised cross sections of forward photon production in DIS
as a function of $x_F$.
The kinematic phase space of the measurements is
given in Table~~\ref{tab:selection}.
For each measurement, the  statistical ($\delta_{stat.}$), 
the total systematic  ($\delta_{total~sys.}$), the
uncorrelated  ($\delta_{uncorrel.sys.}$)
systematic uncertainties and the bin-to-bin correlated
systematic uncertainties due to the FNC absolute 
energy scale  ($\delta_{E_{FNC}}$), 
the measurement of the particle impact position 
in the FNC  ($\delta_{XY_{FNC}}$)
and the model dependence of the data  
correction ($\delta_{model}$) are given.
}
\label{tab:table3}
\end{table}

\begin{table}[hp]
\centering
{
\small
\begin{tabular}{|c|l|l|l||l|l|l|l|}
\multicolumn{8}{c}{$W=70-130~\rm GeV$} \\
\hline  
& & & & & \multicolumn{3}{|c|}{\underline{\hspace*{0.1cm} correlated sys. uncertainty\hspace*{0.1cm}} } \\
$x_F$ range  &  {\Large$\frac{1}{\sigma_{DIS}}\frac{d\sigma}{d x_F}$} & 
$\delta_{stat.}$ & $\delta_{total~sys.}$ &  $\delta_{uncorrel.sys.}$  &  
$\delta_{E_{FNC}}$ & $\delta_{XY_{FNC}}$ & $\delta_{model}$   \\
& & & & &  & & \\ [-5pt]
\hline 
&  & & & & & & \\ [-12pt]
$ 0.10 \div 0.22 $ &  \hspace{5mm}  $0.0456$ & $0.0015$ & $0.0042$  & $0.0012$ & $0.0023$ & $0.0023$  & $0.0023$   \\ 
$ 0.22 \div 0.34 $ &  \hspace{5mm}  $0.0823$ & $0.0016$ & $0.0079$  & $0.0024$ & $0.0049$ & $0.0044$  & $0.0037$  \\
$ 0.34 \div 0.46 $ &  \hspace{5mm}  $0.1096$ & $0.0016$ & $0.0114$  & $0.0033$ & $0.0077$ & $0.0064$  & $0.0044$  \\
$ 0.46 \div 0.58 $ &  \hspace{5mm}  $0.1309$ & $0.0016$ & $0.0151$  & $0.0056$ & $0.0105$ & $0.0076$  & $0.0053$  \\
$ 0.58 \div 0.70 $ &  \hspace{5mm}  $0.1407$ & $0.0015$ & $0.0199$  & $0.0108$ & $0.0127$ & $0.0088$  & $0.0063$  \\ 
$ 0.70 \div 0.82 $ &  \hspace{5mm}  $0.1266$ & $0.0013$ & $0.0179$  & $0.0069$ & $0.0127$ & $0.0085$  & $0.0063$  \\ 
$ 0.82 \div 0.94 $ &  \hspace{5mm}  $0.0656$ & $0.0008$ & $0.0096$  & $0.0036$ & $0.0066$ & $0.0050$  & $0.0033$  \\ 
\hline
\end{tabular}
\vspace*{4mm}

\begin{tabular}{|c|l|l|l||l|l|l|l|}
\multicolumn{8}{c}{$W=130-190~\rm GeV$} \\
\hline  
& & & & & \multicolumn{3}{|c|}{\underline{\hspace*{0.1cm} correlated sys. uncertainty\hspace*{0.1cm}} } \\
$x_F$ range  &  {\Large$\frac{1}{\sigma_{DIS}}\frac{d\sigma}{d x_F}$} & 
$\delta_{stat.}$ & $\delta_{total~sys.}$ &  $\delta_{uncorrel.sys.}$  & 
$\delta_{E_{FNC}}$ & $\delta_{XY_{FNC}}$ & $\delta_{model}$   \\
& & & & & & &  \\ [-12pt]
& & & & & & &  \\ [-8pt]
\hline
& & &  & & & & \\ [-12pt]
$ 0.10 \div 0.22 $ &  \hspace{5mm}  $0.0426$ & $0.0017$ & $0.0038$  & $0.0010$ & $0.0021$ & $0.0021$  & $0.0021$   \\ 
$ 0.22 \div 0.34 $ &  \hspace{5mm}  $0.0801$ & $0.0019$ & $0.0077$  & $0.0023$ & $0.0048$ & $0.0043$  & $0.0036$  \\
$ 0.34 \div 0.46 $ &  \hspace{5mm}  $0.1077$ & $0.0019$ & $0.0112$  & $0.0032$ & $0.0075$ & $0.0063$  & $0.0043$  \\
$ 0.46 \div 0.58 $ &  \hspace{5mm}  $0.1286$ & $0.0018$ & $0.0148$  & $0.0055$ & $0.0103$ & $0.0075$  & $0.0051$   \\
$ 0.58 \div 0.70 $ &  \hspace{5mm}  $0.1359$ & $0.0017$ & $0.0192$  & $0.0105$ & $0.0122$ & $0.0085$  & $0.0061$  \\ 
$ 0.70 \div 0.82 $ &  \hspace{5mm}  $0.1224$ & $0.0014$ & $0.0172$  & $0.0066$ & $0.0122$ & $0.0082$  & $0.0061$  \\ 
$ 0.82 \div 0.94 $ &  \hspace{5mm}  $0.0617$ & $0.0009$ & $0.0090$  & $0.0033$ & $0.0062$ & $0.0047$  & $0.0031$  \\ 
\hline
\end{tabular}
\vspace*{4mm}

\begin{tabular}{|c|l|l|l||l|l|l|l|}
\multicolumn{8}{c}{$W=190-245~\rm GeV$} \\
\hline
& & & & & \multicolumn{3}{|c|}{\underline{\hspace*{0.1cm} correlated sys. uncertainty\hspace*{0.1cm}} } \\
$x_F$ range  &  {\Large$\frac{1}{\sigma_{DIS}}\frac{d\sigma}{d x_F}$} & 
q$\delta_{stat.}$ & $\delta_{total~sys.}$ &  $\delta_{uncorrel.sys.}$  & 
$\delta_{E_{FNC}}$ & $\delta_{XY_{FNC}}$ & $\delta_{model}$  \\
& & & & & & &   \\ [-8pt]
\hline 
&  & & & & & & \\ [-12pt]
$ 0.10 \div 0.22 $ &  \hspace{5mm}  $0.0454$ & $0.0022$ & $0.0042$  & $0.0012$ & $0.0023$ & $0.0023$  & $0.0023$   \\ 
$ 0.22 \div 0.34 $ &  \hspace{5mm}  $0.0796$ & $0.0024$ & $0.0077$  & $0.0023$ & $0.0048$ & $0.0043$  & $0.0036$  \\
$ 0.34 \div 0.46 $ &  \hspace{5mm}  $0.1093$ & $0.0024$ & $0.0114$  & $0.0033$ & $0.0077$ & $0.0064$  & $0.0044$  \\
$ 0.46 \div 0.58 $ &  \hspace{5mm}  $0.1273$ & $0.0023$ & $0.0146$  & $0.0054$ & $0.0102$ & $0.0074$  & $0.0051$   \\
$ 0.58 \div 0.70 $ &  \hspace{5mm}  $0.1357$ & $0.0021$ & $0.0191$  & $0.0104$ & $0.0122$ & $0.0085$  & $0.0061$  \\ 
$ 0.70 \div 0.82 $ &  \hspace{5mm}  $0.1250$ & $0.0018$ & $0.0176$  & $0.0067$ & $0.0125$ & $0.0084$  & $0.0062$  \\ 
$ 0.82 \div 0.94 $ &  \hspace{5mm}  $0.0621$ & $0.0011$ & $0.0090$  & $0.0033$ & $0.0062$ & $0.0047$  & $0.0031$  \\ 
\hline
\end{tabular}
\normalsize

}
\caption{
Normalised cross sections of forward neutron production in DIS
as a function of $x_F$.
The kinematic phase space of the measurements is
given in Table~~\ref{tab:selection}.
For each measurement, the  statistical ($\delta_{stat.}$), 
the total systematic  ($\delta_{total~sys.}$), the
uncorrelated  ($\delta_{uncorrel.sys.}$)
systematic uncertainties and the bin-to-bin correlated
systematic uncertainties due to the FNC absolute 
energy scale  ($\delta_{E_{FNC}}$), 
the measurement of the particle impact position 
in the FNC  ($\delta_{XY_{FNC}}$)
and the model dependence of the data 
correction ($\delta_{model}$) are given.
}
\label{tab:table4}
\end{table}

\newpage

\vspace*{92mm}

\vspace*{80mm}

\newpage

\newpage

\end{document}